\documentclass{emulateapj}

\usepackage{rotating}
\usepackage{natbib}
\usepackage{apjfonts}
\newcommand\tabspace{\noalign{\vspace*{0.7mm}}}
\def\errtwo#1#2#3{$#1^{+#2}_{-#3}$}

\newcommand\msun{{\rm M_\odot}}

\newcommand\fu{{4U~1957+11}}

\newcommand\lmcxt{{LMC~X-3}}
\newcommand\lmcxo{{LMC~X-1}}

\newcommand\zth{$0^{\rm th}$}
\newcommand\fst{$1^{\rm st}$}
\newcommand\scnd{$2^{\rm nd}$}
\newcommand\trd{$3^{\rm rd}$}

\newcommand\arf{{\tt arf}}
\newcommand\arfs{{\tt arfs}}
\newcommand\asca{{ASCA}}
\newcommand\asm{{ASM}}

\newcommand\chandra{{Chandra}}

\newcommand\heg{{HEG}}
\newcommand\hetg{{HETG}}
\newcommand\hexte{{HEXTE}}
\newcommand\isis{{\tt ISIS}}
\newcommand\meg{{MEG}}

\newcommand\pca{{PCA}}
\newcommand\rmf{{\tt rmf}}
\newcommand\rmfs{{\tt rmf}}
\newcommand\rxte{{RXTE}}

\newcommand\xmm{{XMM-Newton}}
\newcommand\rgs{{RGS}}
\newcommand\rgsa{{RGS1}}
\newcommand\rgsb{{RGS2}}

\newcommand\aproxgt{\mathrel{%
      \rlap{\raise 0.511ex \hbox{$>$}}{\lower 0.511ex \hbox{$\sim$}}}}
\newcommand\aproxlt{\mathrel{%
      \rlap{\raise 0.511ex \hbox{$<$}}{\lower 0.511ex \hbox{$\sim$}}}}

\slugcomment{Accepted 2008, July}

\shorttitle{Disk Dominated States of \fu}
\shortauthors{Nowak et al.}

\begin{document}

\title{Disk Dominated States of \fu: {\sl Chandra}, {\sl XMM}, and {\sl RXTE} 
Observations \\ of Ostensibly the Most Rapidly Spinning Galactic Black Hole}

\author{Michael A. Nowak\altaffilmark{1}, Adrienne
  Juett\altaffilmark{2}, Jeroen Homan\altaffilmark{1}, Yangsen
  Yao\altaffilmark{1}, \\ J\"orn Wilms\altaffilmark{3}, Norbert
  S. Schulz\altaffilmark{1}, Claude R. Canizares\altaffilmark{1}}
\altaffiltext{1}{Massachusetts Institute of Technology, Kavli
  Institute for Astrophysics, Cambridge, MA 02139, USA; mnowak,
  jeroen,yaoys,nss,crc@space.mit.edu} \altaffiltext{2}{University of
  Virginia, Dept. of Astronomy, P.O. Box 400325, Charlottesville, VA
  22904-4325, USA; current affiliation: NASA Postdoctoral Fellow,
  Laboratory for X-ray Astrophysics, NASA Goddard Space Flight Center,
  Greenbelt, MD 20771; Adrienne.M.Juett@nasa.gov}
\altaffiltext{3}{Dr. Karl Remeis-Sternwarte, Astronomusches Institut
  der Universit\"at Erlangen-N\"urnberg, Sternwartstr. 7, 96049
  Bamberg, Germany; joern.wilms@sternwarte.uni-erlangen.de}

\begin{abstract}
We present simultaneous \chandra-{High Energy Transmission Gratings}
(\hetg) and {Rossi X-ray Timing Explorer} (\rxte) observations of a
moderate flux `soft state' of the black hole candidate \fu.  These
spectra, having a minimally discernible hard X-ray excess, are an
excellent test of modern disk atmosphere models that include the
effects of black hole spin.  The \hetg\ data show, by modeling the
broadband continuum and direct fitting of absorption edges, that the
soft disk spectrum is only very mildly absorbed with ${\rm N_H} =
1$--$2 \times 10^{21}~{\rm cm}^{-2}$.  These data additionally reveal
$\lambda\lambda13.449$ \ion{Ne}{9} absorption consistent with the
warm/hot phase of the interstellar medium.  The fitted disk model
implies a highly inclined disk around a low mass black hole rapidly
rotating with normalized spin $a^* \approx 1$.  We show, however, that
pure Schwarzschild black hole models describe the data extremely well,
albeit with large disk atmosphere ``color-correction'' factors.
Standard color-correction factors can be attained if one additionally
incorporates mild Comptonization. We find that the
\chandra\ observations do not uniquely determine spin, even with this
otherwise extremely well-measured, nearly pure disk
spectrum. Similarly, {XMM-Newton}/\rxte\ observations, taken only six
weeks later, are equally unconstraining.  This lack of constraint is
partly driven by the unknown mass and unknown distance of \fu;
however, it is also driven by the limited bandpass of \chandra\ and
{XMM-Newton}.  We therefore present a series of 48 \rxte\ observations
taken over the span of several years and at different
brightness/hardness levels.  These data prefer a spin of $a^* \approx
1$, even when including a mild Comptonization component; however, they
also show evolution of the disk atmosphere color-correction factors.
If the rapid spin models with standard atmosphere color-correction
factors of ${\rm h_d} = 1.7$ are to be believed, then the
\rxte\ observations predict that \fu\ can range from a 3\,$\msun$
black hole at 10\,kpc with $a^*\approx 0.83$ to a 16\,$\msun$ black
hole at 22\,kpc with $a^* \approx 1$, with the latter being
statistically preferred at high formal significance.
\end{abstract}


\section{Introduction}\label{sec:intro}

\setcounter{footnote}{0}

Despite having an X-ray brightness comparable to and occasionally
exceeding that of \lmcxo\ or \lmcxt, and despite also being one of the
few persistent black hole candidates (BHC), \fu\ has received
relatively little attention. Similar to \lmcxt\ \citep[see,
e.g.,][]{wilms:01a}, \fu\ is usually in a soft state that shows long
term (hundreds of days) variations \citep{nowak:99d,wijnands:02c}.
Long term variations also have been seen in its optical lightcurve:
modulation over its 9.33\,hr orbital period has ranged from $\pm 10\%$
and sinusoidal \citep{thorstensen:87a} to $\pm 30\%$ and complex
\citep{hakala:99a}.  \citet{hakala:99a} interpret these changes in the
optical lightcurve as evidence for an accretion disk with a large
outer rim, possibly due to a warp, being nearly edge-on and partly
occulted by the secondary.  The lack of any X-ray evidence for 
binary orbital modulation \citep{nowak:99d,wijnands:02c} indicates
that the orbital inclination cannot exceed $\approx 75^\circ$.

Observations with {Ginga} showed a soft spectrum that could be
fit with a multi-temperature disk blackbody \citep[{\tt
    diskbb};][]{mitsuda:84a} plus a power law tail with photon index
$\Gamma \approx 2$--3 \citep{yaqoob:93a}.  The power-law component in
those observations comprised up to 25\% of the 1-18\,keV flux.  Later
\rxte\ observations also were consistent with a disk blackbody model,
but showed no evidence of either a hard component or any X-ray
variability above background fluctuations \citep{nowak:99d}.
\citet{wijnands:02c} conducted a series of Target of Opportunity
observations designed to catch \fu\ at the high end of its count rate
as determined by the {All Sky Monitor} (\asm) on-board of
\rxte.  They found that at its highest flux, \fu\ exhibited both a
hard tail and mild X-ray variability.

For all of the above cited X-ray observations, the disk parameters had
two attributes in common: the best fit inner disk temperatures were
high (up to 1.7\,keV), and the fitted normalizations were low ($\sim
10$).  Nominally, the disk blackbody normalization corresponds to
$(R_{\rm km}/D_{\rm 10})^2 \cos\theta$, where $R_{\rm km}$ is
the disk inner radius in {\rm km}, $D_{\rm 10}$ is the source
distance in units of 10\,kpc, and $\theta$ is the inclination of the
disk.  In physical models, such large temperatures with such low
normalizations normally can be achieved only by a combination of large
distance, high inclination, low black hole mass, high accretion rate,
and possibly rapid black hole spin.  Disk temperatures increase as the
1/4 power of fractional Eddington luminosity, but decrease as the 1/4
power of black hole mass.  Thus, lower mass serves to increase the
temperature and decrease the normalization via decreasing $R_{\rm
km}$.  Large distances and high inclination also serve to reduce
the normalization.  (Again, the lack of X-ray eclipses limits
$i\aproxlt 75^\circ$.)  Finally, rapid spin yields higher efficiency
in the accretion disk (and thus higher temperatures), and a decreased
inner disk radius.

As pointed out by a number of authors, however, the {\tt diskbb} model
is not a self-consistent description of a physical accretion disk
\citep[see, for example,][]{li:05a,davis:05a,davis:06a}.  Several
authors have developed more sophisticated models that incorporate a
torque (or lack thereof) on the inner edge of the disk, atmospheric
effects on the emerging disk radiation field, the effects of limb
darkening and returning radiation (due to gravitational light
bending), and the effects of black hole spin.  Two of these models are
the {\tt kerrbb} model of \citet{li:05a} and the {\tt bhspec} model of
\citet{davis:05a}.  Using these models on soft-state BHC spectra with
minimal hard tails, various claims have been made as to their ability
to observationally constrain black hole spin
\citep[e.g.,][]{shafee:06a,davis:06a,mcclintock:06a,middleton:06a}.
Given the apparent dominance of a soft spectrum in \fu, and the
typical weakness of any hard tail, \fu\ becomes an excellent testbed
for assessing the ability of these models to constrain physical
parameters.

The outline of this paper is as follows.  First, we describe the
analysis procedure for the \chandra, {XMM-Newton}, and \rxte\ data
(\S\ref{sec:data}).  We then discuss fits to the \chandra\ data with
both phenomenological (i.e., {\tt diskbb}) and more physically
motivated (i.e., {\tt kerrbb}) models (\S\ref{sec:broad}).  We then
discuss the {XMM-Newton} data, and specifically look at the absorption
edge regions in both the \chandra\ gratings spectra and the
{XMM-Reflection Gratings Spectrometer} ({RGS}) spectra
(\S\ref{sec:edge}).  We then discuss spectral fits to 48 observations
from the \rxte\ archives (\S\ref{sec:rxte_spectra}), and we consider
the variability properties of these observations
(\S\ref{sec:rxte_vary}).  Finally, we present our conclusions and
summary (\S\ref{sec:discuss}).

\section{Data Preparation}\label{sec:data}

\subsection{Chandra}\label{subsec:chandra}

\fu\ was observed by \chandra\ on 2004 Sept. 7 (ObsID 4552) for
67\,ksec (two binary orbital periods).  The {High Energy
Transmission Gratings} \citep[\hetg;][]{canizares:05a} were inserted.
The \hetg\ is comprised of the {High Energy Gratings} (\heg),
with coverage from $\approx 0.7$--8\,keV, and the {Medium
Energy Gratings} (\meg), with coverage from $\approx 0.4$--8\,keV.
The data readout mode was Timed Exposure-Faint.  To minimize pileup in
the gratings spectra, a 1/2 subarray was applied to the CCDs, and the
aimpoint of the gratings was placed closer to the CCD readout. This
configuration reduces the readout time to 1.741\,sec, without any loss
of the dispersed spectrum.

We used {\tt CIAO v3.3} and {\tt CALDB v3.2.0} to extract the data and
create the spectral response files\footnote{We did, however, use a
  pre-release version of the {\tt Order Sorting and Integrated
    Probability} ({\tt OSIP}) file from CALDB v3.3.0.}. The location
of the center of the \zth\ order image was determined using the {\tt
  findzo.sl} routine\footnote{{\tt
    http://space.mit.edu/ASC/analysis/findzo/}}, which provides
$\approx 0.1$ pixel ($\approx 0.001$\,\AA, for \meg) accuracy.  The
data were reprocessed with pixel randomization turned off, but pha
randomization left on.  We applied the standard grade and bad pixel
file filters, but we did not destreak the data.  (We find that for
sources as bright as \fu, destreaking the data can remove real photon
events, while order sorting is already very efficient at removing
streak events.)

Although our instrumental set up was designed to minimize pileup, it
is still present in both the \meg\ and \heg\ spectra.  Pileup serves
to reduce the peak of the \meg\ spectra, which occurs near 2\,keV, by
approximately 13\%, while it reduces the peak of the \heg\ spectra by
approximately 3\%.  In the Appendix we describe how we incorporate the
effects of pileup in our model fits to the \chandra\ data.

All analyses, figures, and tables presented in this work were produced
using the {\tt Interactive Spectral Interpretation System} (\isis;
\citealt{houck:00a}).  (Note: all `unfolded spectra' shown in the
figures were generated using the model-independent definition provided
by \isis; see \citealt{nowak:05a} for further details.)  For our
$\chi^2$-minimization fits to \chandra\ data, we take the statistical
variance to be the predicted model counts, as opposed to the observed
data counts.  (Choosing the former aids in our Bayesian line searchs;
see \S\ref{sec:edge}.) Additionally, we evaluated the model for each
individual gratings arm, but calculated the fit statistic using the
\isis\ {\tt combine\_datasets} function to add the data from the
gratings arms\footnote{Owing to the fact that the pileup correction is
  a convolution model that requires knowledge of the individual
  response functions for each gratings arm, the dataset combination
  performed here is distinct from adding the PHA and response files
  before fitting. (As a result, this combined data analysis is, in
  fact, impossible to reproduce in {\tt XSPEC}.)}.  The \chandra\ data
show very uniform count rates and spectral colors over the course of
the observation, therefore throughout we use data from the entire
67\,ksec observation.

\subsection{XMM-Newton}\label{subsec:xmm}

\fu\ was observed by \xmm\ on 2004 Oct. 16 (ObsID 206320101) for
45\,ksec.  \xmm\ carries three different instruments, the European
Photon Imaging Cameras
\citep[{EPIC};][]{struder:01a,turner:01a} the Reflection
Grating Spectrometers \citep[{RGS};][]{herder:01a}, and the
Optical Monitor \citep[{OM};][]{mason:01a}.  The {OM}
was not used in this analysis.

The {EPIC} instruments consist of 3 CCD cameras, {MOS-1}, {MOS-2}, and
{pn}, each of which provides imaging, spectral and timing data.  The
{pn} and {MOS-1} cameras were run in timing mode which provides high
time resolution event information by sacrificing 1-dimension in
positional information.  The {MOS-2} camera was run in full-frame
mode.  The {EPIC} instruments provide good spectral resolution
($\Delta E$$=$50--200\,eV FWHM) over the 0.3--12.0 keV range.  There
are two {RGS} detectors onboard \xmm\ which provide high-resolution
spectra ($\lambda/\Delta\lambda$$=$100--500 FWHM) over the
5--38\,\AA\, range.  The grating spectra are imaged onto CCD cameras
similar to the {EPIC-MOS} cameras which allows for order sorting of
the high-resolution spectra.

The \xmm\ data were reduced using {\tt SAS version 7.0}.  Standard
filters were applied to all \xmm\ data.  The {EPIC-pn} data
were reduced using the procedure {\tt epchain}.  We reduced the
{EPIC-MOS} data using {\tt emchain}.  Source and background
spectra for the {pn} and {MOS-1} data were extracted
using filters in the one spatial coordinate, {\tt RAWX}. Response
files were created using the {\tt SAS} tools {\tt rmfgen} and {\tt
  arfgen}.  The {MOS-2} data were found to be considerably
piled up and were therefore not used in this analysis.

The {RGS} data were reduced using {\tt rgsproc}, which produced
standard first order source and background spectra and response files
for both detectors.

%%%% Fig. 1 %%%%

\begin{figure}
\epsscale{1}
\plotone{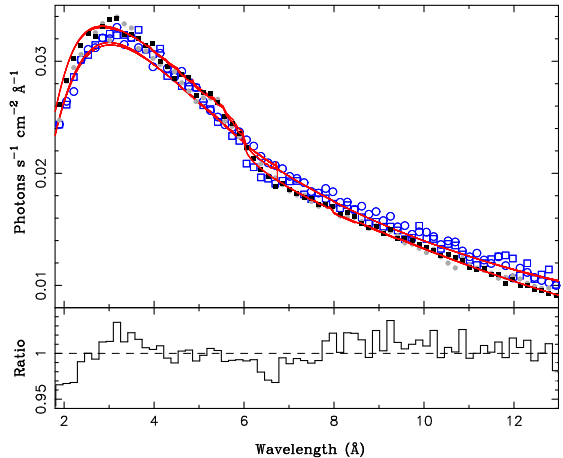}
\caption{Fits to the summed \heg\ (hollow points) and \meg\ (solid
  points) data in the 1.8--13\,\AA range.  The neutral column was
  fixed to be the same for all data sets, but pileup correction was
  treated on an individual detector basis, and the \heg\ {\tt diskbb}
  parameters were allowed to differ from the \meg\ {\tt diskbb}
  parameters.
\label{fig:combo}}
\end{figure}  

\subsection{RXTE}\label{subsec:rxte}

During the past ten years, \rxte\ has observed \fu\ numerous times.
The first of these observations was presented in \citet{nowak:99d}.  A
series of observations, specifically triggered to observe high flux
states, were presented in \citet{wijnands:02c}. One \rxte\ observation
was scheduled to occur simultaneously with the \chandra\ observation
(PI: Nowak), while another was scheduled to occur simultaneously with
the \xmm\ observation (PI: Homan).  The remaining observations come
from different monitoring campaigns\footnote{Some of these
observations were scheduled simultaneously with ground-based optical
observations.  No optical data are included in this work.}, conducted
by various groups.  We obtained all of these data from the archives --
108 ObsIDs, 4 of which we exclude as they occurred during a brief
period when \rxte\ experienced poor attitude control.  We combine
ObsIDs wherein the observations were performed within a few days of
one another and color-intensity diagrams indicate little or no
evolution of the source spectra, yielding 48 spectra.

The data were prepared with the tools from the HEASOFT v6.0 package.
We used standard filtering criteria for data from the {\sl
Proportional Counter Array} (\pca).  Specifically, we excluded data
from within 30 minutes of South Atlantic Anomaly (SAA) passage, from
whenever the target elevation above the limb of the earth was less
than $10^\circ$, and from whenever the electron ratio (a measure of
the charged particle background) was greater than 0.15.  Owing to the
very modest flux of \fu, we used the background models appropriate to
faint data.  We also extracted data from the {\sl High Energy X-ray
Timing Experiment} (\hexte), but as \fu\ is both faint and very soft,
none of these data were of sufficient quality to use for further
analysis.

We applied 0.5\% systematic errors to all \pca\ channels, added in
quadrature to the errors calculated from the data count rate.  For all
\pca\ fits, we grouped the data, starting at $\ge 3$\,keV, with the
criteria that the signal-to-noise (after background subtraction, but
excluding systematic errors) in each bin had to be $\ge 4.5$.  We then
only considered data for which the lower boundary of the energy bin
was $\ge 3$\,keV, and the upper boundary of the energy bin was $\le
18$\,keV.  This last criterion yielded upper cutoffs ranging from
12--18\,keV.

\section{Spectra Viewed with Chandra and XMM-Newton}\label{sec:broad}
\subsection{Broadband Spectra}

Our main goals in modeling the \chandra\ data are to describe the
broad band continuum, accurately fit the edge structure due to
interstellar and/or local-system absorption, and to search for narrow
emission and absorption line features.  To accurately describe the
absorption due to the interstellar medium, we use {\tt tbnew}, an
updated version of the absorption model of \citet{wilms:00a}, which
models the Fe $L_2$ and $L_3$ edges, and includes narrow resonance
line structure in the Ne and O edges, as have been observed at high
spectral resolution with \chandra-\hetg\ \citep[][abundances and
  depletion factors used here have been set to be consistent with
  these previous studies, i.e., they take on the default parameter
  values of the model, with the exception of the Fe abundance
  parameter being set to 0.647, and the Fe and O depletion parameters
  both being set to 1]{juett:04a,juett:06a}.

%%%%% Fig. 2 %%%%

\begin{figure}
\epsscale{1}
\plotone{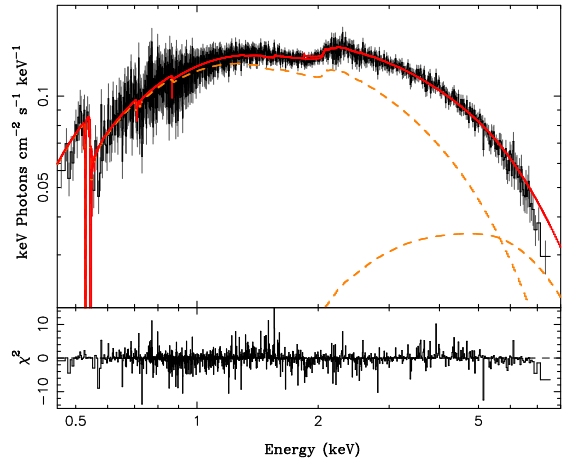}
\caption{A disk plus Comptonization model (see text) fit to the
  combined \meg\ data.  Dashed lines show the individual model
  components.
\label{fig:comptt}}
\end{figure}

We began by jointly modeling both the \heg\ and the \meg\ spectra;
however, as seen in Fig.~\ref{fig:combo}, we find systematic
differences (up to $\approx 5$\%, which is less than the statistical
noise level shown in the figure) between the gratings arms, especially
at low energies, which cannot be accounted for via the pileup
modeling.  Specifically, the \heg\ spectra show a greater soft excess,
such that fits to solely those data in the 0.7--8\,keV range with an
absorbed {\tt diskbb} model yield no absorption, contrary to the edge
structure detected by \meg\ below 0.7\,keV.  Given the disagreements
with \meg, the lack of a fittable column in \heg, the lack of apparent
lines in the 1--8\,keV region (see below), and the fact that the
\meg\ allows us to model line and edge structure in the 0.5--1\,keV
region, we shall only consider \meg\ data.

%%%%% Fig. 3 %%%%

\begin{figure*}
\epsscale{1}
\plottwo{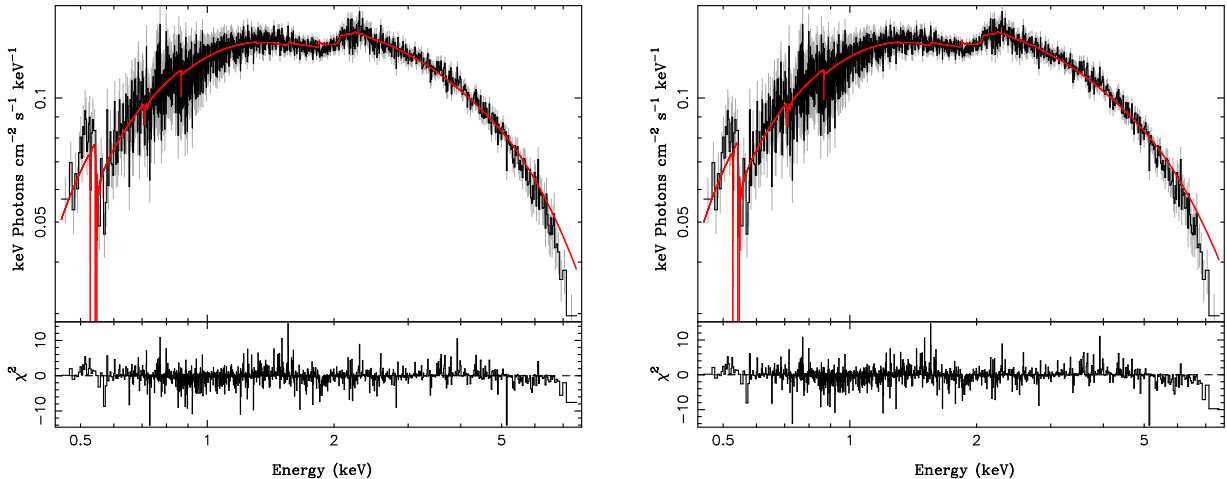}{f3b_col.ps}
\caption{Left: The {\tt kerrbb} model, with $a^*\approx1$, fit to the
  combined \meg\ data.  System distance, inclination, and mass were
  fixed (see text), but spectral hardening factor and black hole spin
  were left free. Right: same data and model as on the left, except
  that now the black hole spin is fixed to $a^* = 0$.
\label{fig:kerrbb}}
\end{figure*}

Systematic disagreements were also found between the simultaneous
\chandra\ and \rxte\ data.  The \rxte\ data were slightly softer in
the regions of overlap ($\approx 4$--7\,keV), and owing to the very
large effective area of \rxte, we found that the latter instrument
dominated any joint fits.  The two instruments fit qualitatively
similar models with comparable parameters; however, no joint model,
even when applying systematic errors to the \pca\ data, produced
formally acceptable fits.  The differences between the instruments are
likely attributable to calibration uncertainties for both, and
therefore we chose to analyze the \chandra\ and \rxte\ data
independently.

For our subsequent broadband spectral fits to the \meg\ data, we
grouped the $\pm1^{\rm st}$ orders to a combined signal-to-noise of
$\ge 6$ and a minimum of two wavelength channels (i.e., 0.01\,\AA) per
bin.  We then fit the data in the 0.45--7.75\,keV region. Results are
presented in Table~\ref{tab:meg_kerrbb}.  Overall, the spectra are
well-described by a very mildly absorbed (${\rm N_H} \approx
1$--$1.2\times 10^{21}~{\rm cm^{-2}}$) {\tt diskbb} spectrum, with a
fairly high temperature (${\rm kT_{in}} = 1.75$\,keV), but low
normalization (${\rm A_{dbb}} =7.2$), with little need for additional
features such as a narrow or broad Fe\,K$\alpha$ line
(cf. \S\ref{sec:edge}).  Formally, if we add a line with fixed energy
and width of 6.4\,keV and 0.5\,keV, respectively, the 90\% confidence
upper limit the line flux is $7\times10^{-5}$ ph\,cm$^{-2}$\,s$^{-1}$.

The low disk model normalization could correspond to a combination of
low black hole mass, high inclination, and large distance.  The fitted
temperature is on the high side for a `soft state' BHC.  Fits to soft
states of LMC~X-1 and LMC~X-3, for example, have consistently found
${\rm kT_{in}} \aproxlt 1.25$\,keV
\citep{nowak:01a,wilms:01a,davis:06a}, while our own fits to \rxte\
and \asca\ data of \fu\ have found ${\rm kT_{in}} \aproxlt 1.6$\,keV.
Such a high temperature might be taken as a hallmark of some
combination of high accretion rate, low black hole mass, and/or high
black hole spin.  Another possibility is the presence of a
Comptonizing corona, serving to harden any disk spectrum.

To show, at least qualitatively, how the presence of a corona would
affect our disk fit parameters, we added the {\tt comptt}
Comptonization model \citep{titarchuk:94a}, to the {\tt diskbb} model,
even though the disk fit residuals do not indicate the necessity of an
extra component.  Given the quality of the simple disk fit, we chose
to reduce the number of free parameters by tying the temperature of
the seed photons input to the corona to the temperature of the disk
inner edge.  Furthermore, we froze the temperature of the corona
itself to 50\,keV.  The additional fit parameters were then the
coronal plasma optical depth, $\tau_{\rm p}$ (with its lower limit set
to 0.01), and the normalization of the coronal spectrum.  Results of
this fit are shown in Fig.~\ref{fig:comptt} and are presented in
Table~\ref{tab:meg_kerrbb}.

\begin{deluxetable*}{ccccccccccc}  
\setlength{\tabcolsep}{0.03in} 
\tabletypesize{\footnotesize}    
\tablewidth{0pt} 
\tablecaption{Broadband Fits to MEG Data \label{tab:meg_kerrbb}}
\tablehead{\colhead{${\rm N_{H}}$}
           & \colhead{${\rm A_{dbb}}$}
           & \colhead{${\rm kT_{in}}$}
           & \colhead{$a^*$}                   
           & \colhead{${\rm \dot M}$}      
           & \colhead{${\rm h_d}$}          
           & \colhead{${\rm P_{frac}}$}
           & \colhead{${\rm A_{comptt}}$}   
           & \colhead{$\tau_{\rm p}$}               
           & \colhead{${\rm kT_0}$}               
           & \colhead{$\chi^2/$dof}         
           \\                               
           $(10^{21}~{\rm cm^{-2}})$
           & & (keV)
           & & ($10^{17}~{\rm g~s^{-1}}$)    
           & & & ($10^{-2}$) & & (keV) 
           }                                  
\startdata
             \errtwo{1.04}{0.04}{0.03} % Nh
           & \errtwo{7.18}{0.18}{0.11}      % Disk norm
           & \errtwo{1.74}{0.02}{0.01} % Disk kT
           & \nodata & \nodata & \nodata
           & \errtwo{0.128}{0.007}{0.008}    % Pile fraction
           & \nodata & \nodata & \nodata
           & 1861/1588                        % chi^2/dof
\\ 
\tabspace  
             \errtwo{1.06}{0.02}{0.04}    % Nh
           & \errtwo{17.87}{1.45}{0.03}   % Disk norm
           & \errtwo{1.25}{0.03}{0.03}    % Disk kT
           & \nodata & \nodata & \nodata
           & \errtwo{0.124}{0.002}{0.008} % Pile fraction
           & \errtwo{0.147}{0.001}{0.002} % Comptt norm
           & \errtwo{0.01}{0.01}{0.00}    % tau_p
           & {\sl 50}                     % Corona kT
           & 1850/1586                    % chi^2/dof
\\ 
\tabspace  
             \errtwo{1.02}{0.02}{0.02} % Nh
           & \nodata & \nodata
           & \errtwo{ 0.9999}{ 0.0000}{ 0.0002} % BH Spin  
           & \errtwo{ 0.548}{ 0.001}{ 0.002} % Mass accretion rate  
           & \errtwo{ 1.220}{ 0.002}{ 0.003} % Hardening factor 
           & \errtwo{0.132}{ 0.002}{ 0.007} % Pileup Fraction
           & \nodata & \nodata & \nodata & 1862/1587 % Chi^2/dof  
\\ 
\tabspace  
             \errtwo{1.07}{0.04}{0.05}               % Nh
           & \nodata & \nodata
           & \errtwo{ 0.952}{ 0.002}{ 0.003}        % BH Spin  
           & \errtwo{ 1.40}{ 0.02}{ 0.03}         % Mass accretion rate  
           & {\sl 1.7}                               % Hardening factor 
           & \errtwo{0.130}{ 0.007}{ 0.008}          % Pileup Fraction
           & \nodata & \nodata & \nodata & 1933/1587 % Chi^2/dof  
\\ 
\tabspace  
             \errtwo{1.13}{0.04}{0.04} % Nh
           & \nodata & \nodata
           & {\sl 0} % BH Spin  
           & \errtwo{ 7.26}{ 0.04}{ 0.04} % Mass accretion rate  
           & \errtwo{ 3.41}{ 0.03}{ 0.02} % Hardening factor 
           & \errtwo{0.126}{ 0.007}{ 0.008} % Pileup Fraction
           & \nodata & \nodata & \nodata & 1892/1588 % Chi^2/dof  
\\ 
\tabspace  
\enddata \tablecomments{The $\chi^2$ statistic was calculated by using
  the predicted model counts as the variance.  Error bars are 90\%
  confidence values ($\Delta \chi^2=2.7$) for one interesting
  parameter. Parameters in italics were held frozen at that value.
  {\tt diskbb} normalization, ${\rm A_{dbb}}$, is ${\rm
    [(R_{in}/1\,km)(D/10\,kpc)]^2\cos\theta}$.  For further details of
  the fits, see the text.}
\end{deluxetable*}

Although the improvement to the fit is only very mild ($\Delta
\chi^2=11$), we see that the temperature of the {\tt diskbb} component
drops to 1.25\,keV, while its normalization constant increases by more
than a factor of 2.  The {\tt comptt} component now comprises $\approx
30\%$ of the implied bolometric flux.  Furthermore, at energies
$\aproxgt 3$\,keV it is consistent with a $\Gamma \approx 2$ photon
index power law; however, this latter fact is tempered by the limited
bandpass of \chandra\ at these energies.  In a ``physical
interpretation'' of these results, we see that a very weak, mild
corona can qualitatively replace the need for an unusually hot, low
normalization disk.

As an alternative for explaining the high disk temperature, we can
consider spin of the black hole.  Although the simple, two parameter
{\tt diskbb} model does an extremely good job of modeling the \meg\
spectra, it has a number of shortcomings from a theoretical
perspective \citep[see][]{davis:06a}.  Its temperature profile is too
peaked at the inner edge, its implied radiative efficiency does not
match realistic expectations, it incorporates no atmospheric physics,
and it includes no (special or general) relativistic effects.  In
response to these model shortcomings, several authors have developed
more sophisticated disk models, specifically the {\tt kerrbb} model of
\citet{li:05a} and the {\tt bhspec} model of \citet{davis:06a}.  Here
we consider the former model.

The {\tt kerrbb} model increases the number of disk parameters from
two to seven, and it allows other effects to be included, such as limb
darkening and returning radiation to the disk from gravitational light
bending (both turned on for fits described in this paper).  The fit
parameters are the mass, spin ($a^*$), and distance to the black hole,
the mass accretion rate through the disk (${\rm \dot M}$), the
inclination of the accretion disk to the line of sight, a torque
applied to the inner edge of the disk (here set to zero), and a
spectral hardening factor (${\rm h_d}$).  The latter is to absorb
uncertainties of the disk atmospheric physics, and represents the
ratio of the disk's color temperature to its effective temperature.
Preferred values have been ${\rm h_d} \sim 1.7$
\citep{li:05a,shafee:06a,mcclintock:06a}, while other models (e.g.,
{\tt bhspec}) attempt to calculate color corrections from first
principles \citep{davis:06a}.

%%%%% Fig. 4 %%%%%

\begin{figure}
\epsscale{1}
\plotone{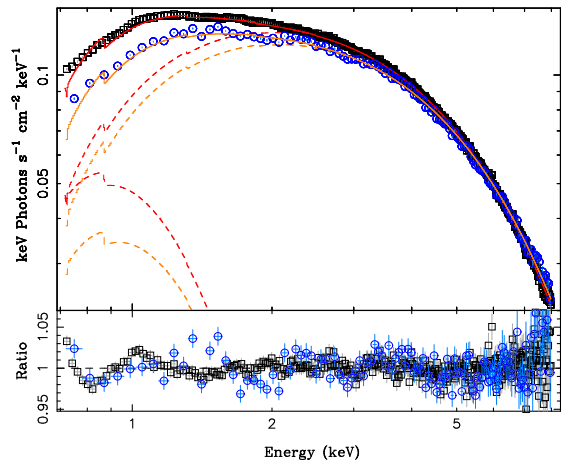}
\caption{Flux-corrected \xmm\ {EPIC-pn} (squares) and {-MOS-1} spectra
  (circles), fit with independent absorbed {\tt diskbb+diskbb} models
  in the 0.7--8\,keV range. Dashed lines show the individual absorbed
  disk components for the fit. 0.05\% systematic errors were added to
  the {pn} data, while 1\% systematic errors were added to the {MOS}
  data.
\label{fig:pnspectra}}
\end{figure}  

To reduce the fit complexity of the {\tt kerrbb} model to match the
two parameters of the {\tt diskbb} model, one typically invokes
knowledge obtained from other observations (i.e., for system mass,
distance, and inclination) and from theoretical expectations (i.e.,
for inner edge torque and ${\rm h_d}$).  The former is lacking for
\fu.  Here we fix the inclination to $75^\circ$, to be consistent with
the both the optical and X-ray lightcurve behavior
\citep{hakala:99a,nowak:99d}.  Furthermore, we fix the mass to
3\,$\msun$ and the distance to 10\,kpc.  We discuss these choices
further in \S\ref{sec:rxte_spectra} and \S\ref{sec:discuss}, but
here we note that these latter choices allow the closest possible
distance to \fu, while still maintaining a minimum inferred bolometric
luminosity of $> 3\%\,{\rm L_{Edd}}$.  Finally, we set the torque
parameter to 0, as its value is essentially subsumed by degeneracies
with the fitted mass accretion rate \citep{li:05a,shafee:06a}.

Allowing the spectral hardening factor, ${\rm h_d}$ to remain free, we
obtain a {\tt kerrbb} fit to the \meg\ spectra that is of comparable
quality to the simple {\tt diskbb} fit.  We note, however, that the
fitted value of ${\rm h_d}=1.22$ is smaller than the commonly
preferred value of 1.7 \citep[see][]{done:08a}, and that the black
hole spin is found to be near unity ($a^* = 0.9999$).  Again, the
latter is being driven by the fact that the fitted {\tt diskbb}
temperature is itself rather high for a soft state BHC system.
Freezing ${\rm h_d}=1.7$, we obtain a slightly worse, but still
reasonable, fit ($\Delta \chi^2=71$), with a large inferred black hole
spin ($a^*=0.95$).  If instead we freeze the spin $a^*=0$, but allow
the spectral hardening factor to be free, we obtain a $\chi^2$
intermediate to those of the previous two fits ($\Delta \chi^2 = 30$
from the best {\tt kerrbb} fit), but with ${\rm h_d}=3.41$. This
rather large value is well-outside the normal range of variation
(${\rm h_d}=$1.4--1.8; \citealt{done:08a}).  As shown in
Fig.~\ref{fig:kerrbb}, however, these fits are virtually
indistinguishable from one another.

We next consider broadband fits to \xmm\ data.  These
\xmm\ observations occured only 6 weeks after the
\chandra\ observations, with apparently little source variation as
measured by the \rxte-\asm\ in the intervening period.  As we further
discuss in \S\ref{sec:rxte}, the \xmm\ spectra showed a flux and
spectral hardness comparable to the \chandra\ spectra.  Thus, we
expect the \xmm\ spectra to be qualitatively and quantitatively
similar to the \chandra\ spectra.  We have found that in practice, the
\xmm\ spectra are somewhat difficult to fit over their entire spectral
range, and furthermore, the {EPIC-pn} and {-MOS} data do
not completely agree with one another.  For a variety of models that
we tried, strong, sharply peaked residuals occurred for both detectors
near 0.5\,keV (the O K edge), although these residuals were most
pronounced in the {pn} data.  Above 8\,keV, the two detectors
also show (different from each other, and from \rxte) spectral
deviations from simple models.  Accordingly, we only consider data in
the 0.7--8\,keV range.  Additionally, we add uniform 0.5\%
({pn}) and 1\% ({MOS}) systematic errors to each
spectrum. (These levels were required to obtain reduced $\chi^2
\approx 1$--2 with `simple' spectral models.)

The only simple models that we found to adequately describe the data
required an additional spectral component below 2\,keV, which here we
model as a \emph{second} {\tt diskbb} component with ${\rm kT_{in}}
\approx 0.3$\,keV and ${\rm A_{dbb}} \approx 1300$ ({MOS}) and
$\approx 2600$ ({pn}).  The required ${\rm N_h}\approx 3 \times
10^{21}\,{\rm cm^{-2}}$ for both detectors was somewhat larger than
that found for \meg.  The dominant disk component, primarily required
to fit the 2-8\,keV spectra, had comparable parameters to the {\tt
  diskbb} fit to the \meg\ data; namely, ${\rm A_{dbb}} \approx 8$ and
${\rm kT_{in}} \approx 1.6$\,keV.

Significant residuals clearly remain in the 0.7--2\,keV range for both
the {pn} and {MOS} data, and the {pn} and
{MOS} spectra quantitatively disagree with each other.  Given
the fact that the \meg\ spectra, as well as prior {ASCA}
spectra \citep{nowak:99d} were well described with a simple absorbed
{\tt diskbb} model, with lower ${\rm N_H}$, we are inclined to ascribe
a large fraction of the additional required $\approx 0.3$\,keV disk
component and the low energy residuals to systematic errors in the
\xmm\ calibration.

With the above caveats in mind, the \xmm\ data, however, confirm that
in the 2--8\,keV range, the characteristic disk temperature for
\fu\ is indeed rather large.  Furthermore, the \xmm\ spectra show no
evidence for any narrow, or moderately broad, Fe K$\alpha$ line.
Similar to the \chandra\ data, adding a 6.4\,keV line with 0.5\,keV
width (both fixed) yields a 90\% confidence upper limit to the line
flux of $6\times10^{-5}$\,ph\,cm$^{-2}$\,s$^{-1}$.

\begin{deluxetable}{ccc}  
\setlength{\tabcolsep}{0.03in} 
\tabletypesize{\footnotesize}    
\tablewidth{0pt} 
\tablecaption{Absorption Lines Fit to MEG and RGS Data \label{tab:lines}}
\tablehead{\colhead{Line}
           & \colhead{$\lambda$}
           & \colhead{Eq. Width}
           \\                               
           & (\AA)
           & (m\AA)
           }                                  
\startdata
             Ne{\sc ix}\tablenotemark{a}
               & \errtwo{13.449}{0.005}{0.005}   % Wavelength
           & \errtwo{-7.2}{2.8}{2.2}      % Equivalent Width
\\ 
\tabspace  
             Ne{\sc iii}
           & {\sl 14.508}   % Wavelength
           & \errtwo{-6.0}{3.3}{3.0}      % Equivalent Width
\\ 
\tabspace  
             Ne{\sc ii}
           & \errtwo{14.611}{0.005}{0.005}   % Wavelength
           & \errtwo{-7.5}{4.8}{3.1}      % Equivalent Width
\\ 
\tabspace  
\hline
\tabspace  
             O{\sc viii} K$\beta$\tablenotemark{b}
           & {\it 18.629}                 % Wavelength
           & \errtwo{-9.9}{9.9}{9.0}      % Equivalent Width
\\ 
\tabspace  
             O{\sc viii} K$\alpha$
           & {\it 18.967}                 % Wavelength
           & \errtwo{-21}{11}{10}         % Equivalent Width
\\ 
\tabspace  
             O{\sc vii}
           & {\it 21.602}                 % Wavelength
           & \errtwo{-18}{18}{16}         % Equivalent Width
\\ 
\tabspace   
             ?
           & \errtwo{21.80}{0.04}{0.04}   % Wavelength
           & \errtwo{-71}{36}{43}         % Equivalent Width
\\ 
\tabspace   
             O{\sc ii}
           & {\it 23.375}                 % Wavelength
           & \errtwo{-15}{15}{18}         % Equivalent Width
\\ 
\tabspace   
\hline
\tabspace  
             O{\sc viii} K$\beta$\tablenotemark{c}
           & {\it 18.629}                 % Wavelength
           & \errtwo{0}{0}{14}            % Equivalent Width
\\ 
\tabspace  
             O{\sc viii} K$\alpha$
           & {\it 18.967}                 % Wavelength
           & \errtwo{-28}{15}{14}         % Equivalent Width
\\ 
\tabspace  
             O{\sc vii}
           & {\it 21.602}                 % Wavelength
           & \errtwo{-36}{29}{24}         % Equivalent Width
\\ 
\tabspace   
             ?
           & \errtwo{21.80}{0.02}{0.02}   % Wavelength
           & \errtwo{-67}{44}{36}         % Equivalent Width
\\ 
\tabspace   
             O{\sc ii}
           & {\it 23.375}                 % Wavelength
           & \errtwo{-11}{11}{29}         % Equivalent Width
\\ 
\tabspace   
\hline
\tabspace  
             O{\sc viii} K$\beta$\tablenotemark{d}
           & \errtwo{18.62}{0.03}{0.03}    % Wavelength
           & \errtwo{-8.5}{8.5}{5.4}       % Equivalent Width
\\ 
\tabspace  
             O{\sc viii} K$\alpha$
           & \errtwo{18.96}{0.02}{0.02}   % Wavelength
           & \errtwo{-14}{14}{5}          % Equivalent Width
\\ 
\tabspace  
             O{\sc vii}
           & \errtwo{21.59}{0.03}{0.03}   % Wavelength
           & \errtwo{-19}{19}{11}         % Equivalent Width
\\ 
\tabspace   
             ?
           & \errtwo{21.83}{0.03}{0.03}   % Wavelength
           & \errtwo{-23}{11}{11}         % Equivalent Width
\\ 
\tabspace   
             O{\sc ii}
           & {\it 23.375}                 % Wavelength
           & \errtwo{-29}{11}{11}         % Equivalent Width
\\ 
\enddata 
\tablenotetext{a}{MEG data fit between 13\,\AA--15\,\AA, grouped to S/N$\ge 5$ per bin.}
\tablenotetext{b}{MEG data fit between 18\,\AA--25\,\AA, grouped to 16 channels per bin.}
\tablenotetext{c}{MEG data fit between 18\,\AA--25\,\AA, grouped to 32 channels per bin.}
\tablenotetext{d}{RGS data fit between 18\,\AA--25\,\AA\ (RGS1) and 18\,\AA-19.9\,\AA\ (RGS2), 
                  grouped to S/N$\ge 5$ and $\ge 4$ channels per bin.}
\tablecomments{Italicized parameters were held fixed.  Errors are 90\% confidence level.}
\end{deluxetable}

\subsection{Edge and line fits}\label{sec:edge}

We have employed a variety of techniques to search for lines in the
\fu\ data, ranging from direct fits of narrow components at known line
locations to a ``blind search'' using a Bayesian Blocks technique.
The latter involves fitting a continuum model to the binned data,
using the predicted model counts as the statistical variance in the
fits, and then unbinning the data and comparing the likelihood of the
observed counts to the predicted counts.  (For a successful
application of this technique to \chandra-\hetg\ data of the low
luminosity active galactic nuclei, M81*, see \citealt{young:07a}.)  No
candidate features that appear to be local to the \fu\ system were
found.  This is in contrast to, for example, \chandra\ observations of
GROJ1655$-$40, which at a similar source luminosity in a spectrally
soft, disk-dominated state showed very strong absorption features
associated with the disk atmosphere \citep{miller:06a}.  Comparable
equivalent width features would have been \emph{very easily} detected
in our observation of \fu.  We note, however, that GROJ1655$-$40
itself has not always shown disk atmosphere absorption features.  Two
weeks prior to the observation described by \citet{miller:06a},
despite the source being only slightly fainter while also being in a
soft, disk-dominated state, \chandra-\hetg\ observations revealed
\emph{no} narrow emission or absorption lines
\citep{miller:06a,miller:08a}.  The `duty cycle' with which such lines
are prominent in disk-dominated BHC spectra comparable to the spectra
of both GROJ1655$-$40 and \fu, remains an open observational question.

%%%%% Fig. 5 %%%%%

\begin{figure}
\epsscale{1} \plotone{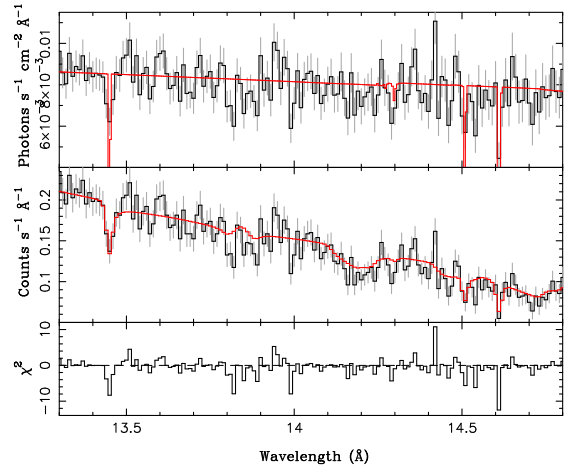}
\caption{Flux corrected spectra (top panel) and count rate per
  angstrom (middle panel) for the combined \meg\ $\pm1^{\rm st}$ order
  spectra, with an absorbed {\tt diskbb} model fit in the
  13--15\,\AA\ region (see text). The top and middle panels show the
  data fit with the \ion{Ne}{2}, \ion{Ne}{3}, and \ion{Ne}{9} lines,
  while the residuals show a fit without the lines.
\label{fig:lines}}
\end{figure}  

On the other hand, the above Bayesian Blocks procedure yielded a few
candidate features in the \chandra\ data that are likely attributable
to absorption by the Interstellar Medium (ISM). The strongest such
feature is the absorption line at 13.44\,\AA. This feature is
consistent with \ion{Ne}{9} absorption from the hot phase of the
interstellar medium \citep{juett:06a}. Accordingly, we also
simultaneously fit expected features from \ion{Ne}{2} and \ion{Ne}{3}
absorption at 14.61\,\AA\ and 14.51\,\AA, respectively.

We again choose an absorbed disk model, albeit with the pileup
fraction fixed to the values from our broad-band models, and only
consider the 13--15\,\AA\ region.  We group the spectrum to a minimum
signal-to-noise of 5 in each bin, and we let the equivalent widths and
wavelengths of the \ion{Ne}{9} and \ion{Ne}{2} lines be free
parameters while fixing the wavelength of the \ion{Ne}{3} absorption
to 14.508\,\AA.  Results from this fit are presented in
Fig.~\ref{fig:lines} and Table~\ref{tab:lines}.  The fitted
wavelengths of the \ion{Ne}{9} and \ion{Ne}{2} lines agree well with
the \meg\ studies of \citet{juett:06a}, although we find the
\ion{Ne}{9} line wavelength to be redshifted by 0.005\,\AA\ (i.e.,
111\,km~s$^{-1}$).  Our fits, however, are within 0.002\,\AA\ of the
theoretical value of 14.447\,\AA.

Formally, the fit improves with addition of any of the three Ne lines,
although the \ion{Ne}{3} region residual is far from the strongest in
the overall spectrum.  The \ion{Ne}{9} and \ion{Ne}{2} absorption lines
are more clearly detected.  Their equivalent widths are near the
mid-level to higher end\footnote{The \ion{Ne}{9} line equivalent width
  measured in \fu\ is a factor of two less than that observed in
  GX~339$-$4; however, as discussed by \citet{miller:04a} and
  \citet{juett:06a} the latter source's \ion{Ne}{9} column is likely
  dominated by a warm absorber intrinsic to the binary.} of the sample
discussed by \citet{juett:06a}.  As we discuss in greater detail
elsewhere \citep{yao:08a}, the \fu\ \ion{Ne}{9} equivalent width is
consistent with the source's probable location outside the disk of the
Galaxy.  Specifically, presuming that \fu\ is greater than 5\,kpc
distant, and choosing a model where the hot phase of the ISM lies in
layers predominantly outside the galactic plane, the equivalent width
of the \ion{Ne}{9} line in \fu\ is consistent with the equivalent width
of the same feature observed in the Active Galactic Nuclei Mkn 421,
appropriately scaled for that sightline through our Galaxy.  Therefore
it is likely that \fu\ is sampling a large fraction of the hot phase
interstellar gas that lies within and directly outside the galactic
plane \citep{yao:07a}.

%%%%%%% Fig. 6 %%%%%

\begin{figure}
\epsscale{1} \plotone{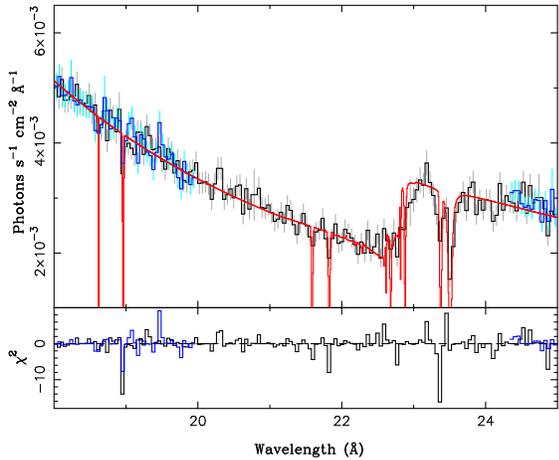}
\caption{Flux corrected \rgs\ spectra in the 18--25\,\AA\ (\rgsa) and
  18-19.9\,\AA\ (\rgsb) range, fit with a model consisting of
  interstellar absorption, a phenomenological continuum model (a
  broken power law plus multicolor blackbody), and various absorption
  lines (see Table~\protect{\ref{tab:lines}}). The residuals show the
  fit without the lines. \label{fig:rgsedge}}
\end{figure}  

Consistent with this picture, both the \rgs\ and \meg\ spectra show
evidence of longer wavelength features that we also associate with
absorption by the interstellar medium (Table~\ref{tab:lines}; see also
\citealt{yao:08a}).  For the \rgs\ data, we grouped both the \rgsa\
(fit between 18--25\,\AA) and \rgsb\ (fit between 18--19.9\,\AA) data
to have a signal-to-noise of 5 and a minimum of four channels per bin.
The \meg\ data were grouped to either 16 channels per bin, or 32
channels per bin.  The \meg\ data were fit with an absorbed disk
blackbody, while the \rgs\ data required additional continuum
structure (which we modeled here with a broken power-law).  For all
these model fits we added absorption lines expected from the warm and
hot phases of the interstellar medium.  The \ion{O}{2} line from the
warm phase of the ISM (not included in the {\tt tbnew} model;
\citealt{juett:04a}) is clearly detected by the \rgs\ data, and
consistent limits are found with the \meg\ data.

The hot phase of the ISM is detected with both the \rgs\ and \meg\ via
\ion{O}{8} K$\alpha$ absorption (18.96\,\AA) measurements.  The
presence of \ion{O}{8} K$\beta$ (18.63\,\AA) is not strictly required,
but the limits on its equivalent width are consistent with the
measured K$\alpha$ line \citep{yao:08a}. In addition to \ion{O}{8}, we
also find evidence for \ion{O}{7} K$\alpha$ absorption.
Interestingly, an even stronger feature is detected at 21.8\,\AA.  We
have no good identification for this feature (it is 3000\,${\rm
  km~s^{-1}}$ redshifted from \ion{O}{7}). Its equivalent width might
change between the \meg\ and \rgs\ observations, and therefore it
could be a candidate for a line intrinsic to the source.

%%%%% Fig. 7 %%%%

\begin{figure}
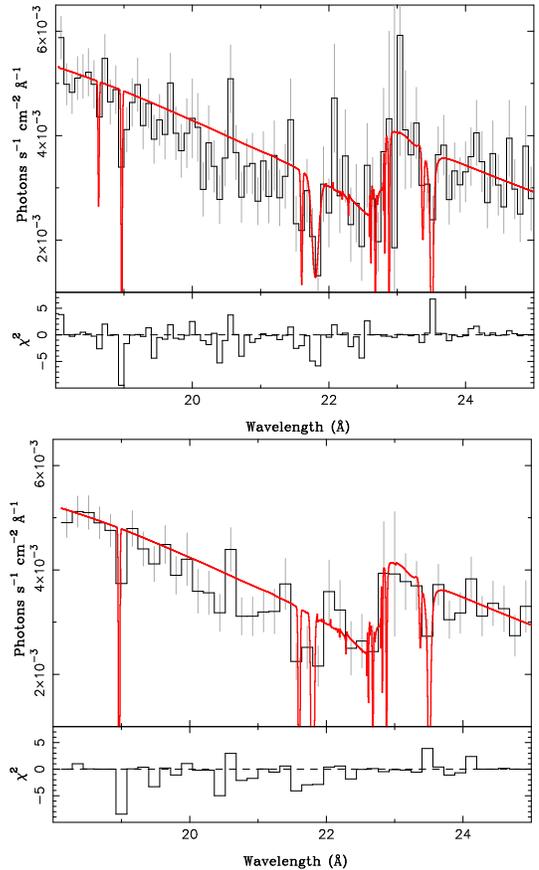

\epsscale{0.95}
\plotone{f7a.ps}
\plotone{f7b.ps}

\caption{Flux corrected \meg\ spectra in the 18--25\,\AA\ range, fit
  with a model consisting of interstellar absorption, a multicolor
  blackbody, and various absorption lines (see
  Table~\protect{\ref{tab:lines}}).  The residuals show the fit
  without the lines.  The top panel shows the spectra grouped to 16
  channels per bin, while the bottom panel shows the spectra grouped
  to 32 channels per bin. \label{fig:megedge}}
\end{figure}  

It is interesting to note that for the 18--25\,\AA\ fits discussed
above, we find that the implied neutral column is somewhat larger than
that found with the broadband fits.  Specifically, we find $N_H =
1.9\pm0.2$, $1.8\pm0.5$, and $2.0\pm0.2$ $\times 10^{21}~{\rm
  cm^{-2}}$ for the \rgs\ and \meg\ (16 channels per bin, 32 channels
per bin) data.  These values are primarily driven by the O edge region
of the fits, although we have found for our broadband fits that merely
changing the O abundance is insufficient to bring their fitted neutral
columns into agreement with the 18--25\,\AA\ fits. It is possible that
an additional very low energy component added to the broad band fits
might allow for a larger column that is consistent with these narrow
band fits.

\section{RXTE Observations}\label{sec:rxte}

In the previous sections we have shown that the \fu\ spectrum is
consistent with a disk having a high peak temperature.  One could
reduce the need for this large peak temperature by including a hard
component, e.g., a Comptonization spectrum; however, this added
component only becomes dominant at photon energies above the
bandpasses of \chandra\ and \xmm.  Thus, to distinguish between the
possibilities of high black hole spin or an additional hard spectral
component leading to the high fitted temperatures, we turn toward the
\rxte\ data, which can provide spectral information up to $\approx
20$\,keV for \fu.

\begin{deluxetable*}{ccccccccccc}  
\setlength{\tabcolsep}{0.03in} 
\tabletypesize{\footnotesize}    
\tablewidth{0pt} 
\tablecaption{{\tt (diskbb+gaussian)+powerlaw}, or {\tt +comptt} Fits to Selected RXTE Data \label{tab:rxtedbb}} 
\tablehead{ \colhead{ID}                    
           & \colhead{MJD}                   
           & \colhead{${\rm A_{dbb}}$}      
           & \colhead{${\rm T_{in}}$}       
           & \colhead{${\rm A_{pl}}$}       
           & \colhead{$\Gamma$}             
           & \colhead{${\rm A_{comptt}}$}   
           & \colhead{$\tau_{\rm p}$}      
           & \colhead{${\rm A_{line}}$}     
           & \colhead{$\sigma_{\rm line}$} 
           & \colhead{$\chi^2/$dof}         
           \\                               
           & &                                
           & (keV)                            
           & $(10^{-2})$                      
           & & $(10^{-5})$                    
           & & $(10^{-4})$                    
           & (keV)                            
           }                                  
\startdata  
40044-01-02-01 % ObsID 
           &    51271 % MJD  
           & \errtwo{  5.05}{  0.47}{  0.47} % Diskbb norm  
           & \errtwo{ 1.801}{ 0.022}{ 0.019} % Diskbb kT  
           & \errtwo{28.54}{ 5.45}{ 4.99} % Powerlaw norm  
           & \errtwo{ 2.13}{ 0.07}{ 0.07} % Gamma  
           & \nodata & \nodata & \errtwo{  3.9}{  1.7}{  1.7} % Line norm  
           & \errtwo{  0.0}{  0.6}{  0.0} % Line sigma  
           & 48.9/28 % Chi^2/dof  
\\ 
\tabspace  
           & & \errtwo{ 24.98}{  0.02}{  3.03} % Diskbb norm  
           & \errtwo{ 1.226}{ 0.001}{ 0.001} % Diskbb kT  
           & \nodata & \nodata &  \errtwo{249.8}{ 0.8}{ 0.8} % Comptt norm  
           & \errtwo{ 0.22}{ 0.00}{ 0.00} % Comptt Optical Depth  
           & \errtwo{  2.4}{  1.2}{  1.2} % Line norm  
           & \errtwo{  0.0}{  0.8}{  0.0} % Line sigma  
           & 25.1/28 % Chi^2/dof  
\\ 
\tabspace  
70054-01-11-00 % ObsID 
           &    52656 % MJD  
           & \errtwo{ 12.40}{  0.05}{  0.04} % Diskbb norm  
           & \errtwo{ 1.313}{ 0.001}{ 0.001} % Diskbb kT  
           & \errtwo{ 0.00}{ 0.45}{ 0.00} % Powerlaw norm  
           & \errtwo{ 3.00}{ 0.00}{ 2.00} % Gamma  
           & \nodata & \nodata & \errtwo{  3.9}{  0.5}{  0.5} % Line norm  
           & \errtwo{  0.8}{  0.2}{  0.1} % Line sigma  
           & 16.7/18 % Chi^2/dof  
\\ 
\tabspace  
           & & \errtwo{ 12.52}{  0.04}{  0.08} % Diskbb norm  
           & \errtwo{ 1.311}{ 0.001}{ 0.001} % Diskbb kT  
           & \nodata & \nodata &  \errtwo{ 0.0}{ 1.1}{ 0.0} % Comptt norm  
           & \errtwo{ 0.53}{ 4.47}{ 0.52} % Comptt Optical Depth  
           & \errtwo{  4.2}{  0.5}{  0.9} % Line norm  
           & \errtwo{  0.8}{  0.2}{  0.2} % Line sigma  
           & 16.4/18 % Chi^2/dof  
\\ 
\tabspace  
90123-01-03 % ObsID 
           &    53255 % MJD  
           & \errtwo{ 11.49}{  0.24}{  0.32} % Diskbb norm  
           & \errtwo{ 1.554}{ 0.008}{ 0.007} % Diskbb kT  
           & \errtwo{ 0.71}{ 4.31}{ 0.70} % Powerlaw norm  
           & \errtwo{ 2.38}{ 0.62}{ 1.38} % Gamma  
           & \nodata & \nodata & \errtwo{  4.6}{  1.9}{  1.5} % Line norm  
           & \errtwo{  0.5}{  0.3}{  0.4} % Line sigma  
           & 25.7/26 % Chi^2/dof  
\\ 
\tabspace  
           & & \errtwo{ 12.10}{  0.03}{  0.72} % Diskbb norm  
           & \errtwo{ 1.533}{ 0.026}{ 0.013} % Diskbb kT  
           & \nodata & \nodata &  \errtwo{ 5.6}{ 0.4}{ 4.8} % Comptt norm  
           & \errtwo{ 0.01}{ 4.99}{ 0.00} % Comptt Optical Depth  
           & \errtwo{  4.9}{  1.8}{  0.8} % Line norm  
           & \errtwo{  0.5}{  0.3}{  0.2} % Line sigma  
           & 25.3/26 % Chi^2/dof  
\\ 
\tabspace  
90063-01-01 % ObsID 
           &    53294 % MJD  
           & \errtwo{ 11.48}{  0.29}{  0.30} % Diskbb norm  
           & \errtwo{ 1.552}{ 0.007}{ 0.006} % Diskbb kT  
           & \errtwo{ 1.61}{ 0.02}{ 1.60} % Powerlaw norm  
           & \errtwo{ 2.60}{ 0.40}{ 1.60} % Gamma  
           & \nodata & \nodata & \errtwo{  5.0}{  1.8}{  1.6} % Line norm  
           & \errtwo{  0.6}{  0.2}{  0.3} % Line sigma  
           & 19.0/26 % Chi^2/dof  
\\ 
\tabspace  
           & & \errtwo{ 12.24}{  0.42}{  0.78} % Diskbb norm  
           & \errtwo{ 1.529}{ 0.029}{ 0.015} % Diskbb kT  
           & \nodata & \nodata &  \errtwo{ 6.4}{ 0.8}{ 5.1} % Comptt norm  
           & \errtwo{ 0.01}{ 4.99}{ 0.00} % Comptt Optical Depth  
           & \errtwo{  5.2}{  1.6}{  1.7} % Line norm  
           & \errtwo{  0.6}{  0.2}{  0.3} % Line sigma  
           & 18.9/26 % Chi^2/dof  
\\ 
\tabspace  
\enddata 
\tablecomments{ {\tt diskbb} normalization, 
${\rm A_{dbb}}$, is ${\rm [(R_{in}/1\,km)(D/10\,kpc)]^2\cos\theta}$. 
Power law normalization, ${\rm A_{pl}}$, is photons/cm$^2$/s/keV at 1\,keV. 
Line normalization, ${\rm A_{line}}$, is total photons/cm$^2$/s in the line. 
0.5\% systematic errors were added in quadrature to the data.  Error bars 
are 90\% confidence level} 
\end{deluxetable*}

\subsection{Spectra}\label{sec:rxte_spectra}

\begin{figure}
\epsscale{1}
\plotone{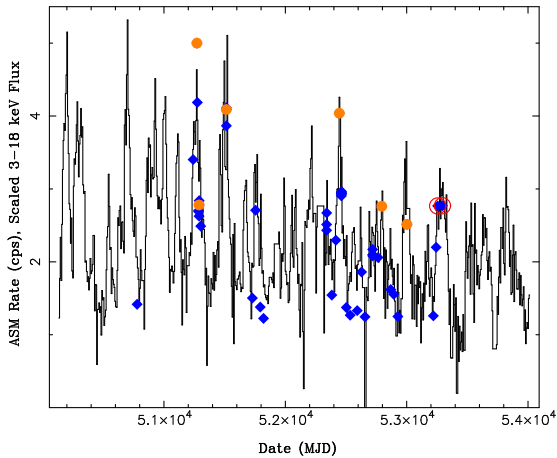}
\caption{{All Sky Monitor} lightcurve for \fu\, with scaled 3-18\,keV
  \pca\ flux from pointed observations overlaid.  \rxte\ observations
  simultaneous with \chandra\ and \xmm\ are circled. Circles here and
  in subsequent figures represent observations where \fu\ was
  transiting to `steep power-law state' behavior.
\label{fig:lightcurve}}
\end{figure}

In Fig.~\ref{fig:lightcurve} we present the \rxte-\asm\ lightcurve,
with 6\,day bins, and also show the scaled 3--18\,keV flux from the
pointed \rxte\ observations.  This figure highlights the times of the
\chandra\ and \xmm\ observations (which were simultaneous with
\rxte\ ObsIDs 90123-01-03 and 90063-01-01, respectively).  Generally
speaking, \fu\ exhibits an X-ray lightcurve with a factor of 3--4
variability on a few hundred day time scale.  \citet{nowak:99d}
hypothesized that this long term variability was quasi-periodic;
however, our longer lightcurve does not show any clear super-orbital
periodicities.  Note that the \chandra\ and \xmm\ observations occur
during the same local peak in the \asm\ lightcurve and have very
similar 3--18\,keV fluxes as measured by the \pca.  We shall present
detailed results for these two observations, as well as for the
faintest \rxte\ observation (ObsID 70054-01-11-00) and the brightest
\rxte\ observation (ObsID 40044-01-02-01).  This latter observation
has been presented previously by \citet{wijnands:02c}.

%%%% Fig. 8 %%%%

As discussed by \citet{nowak:99d} and \citet{wijnands:02c}, the
\rxte\ data of \fu\ can be well-described by a model that consists of
a combination of a multi-temperature disk spectrum, a power-law, and a
gaussian line.  For most of the \rxte\ observations the line peaks at
an amplitude of $\aproxlt 2$--$3\%$ of the local continuum level, and
it is statistically required by the data.  No such line was present in
either the \chandra\ or the \xmm\ observation (line flux $\aproxlt
7\times10^{-5}$\,ph\,cm$^{-2}$\,s$^{-1}$). There are several
possibilities for this discrepancy, and perhaps some combination of
all are at play.  With its $\approx 1\,{\rm deg}^2$ effective field of
view, the \pca\ will detect diffuse galactic emission which shows a
prominent 6.7\,keV line.  Based upon scalings between the \pca\ line
flux and the DIRBE 4.9\,$\mu$m infrared flux \citep{revnivtsev:06a},
one expects an \rxte-detected line amplitude of $\approx
4\times10^{-5}$\,ph\,cm$^{-2}$\,s$^{-1}$ (i.e., approximately 10\% of
the observed line) at the Galactic coordinates of
\fu\ ($l=51.307^\circ$, $b=-9.330^\circ$).  Remaining systematic
uncertainties in the \pca\ response matrix may be as large as 1\% of
the continuum level in the Fe line region, which could account for a
line amplitude of $\aproxlt 1.5\times10^{-4}$\,ph\,cm$^{-2}$\,s$^{-1}$
for the observations simultaneous with \chandra\ and \xmm.  The
remaining 50\% of the line flux for these latter two observations,
which is still three times larger than the \chandra\ and \xmm\ upper
limits, may be real and could be indicative of the greater sensitivity
of \pca\ to weak and \emph{broad} line features.  Such a broad line
would have been detectable by \chandra\ and \xmm\ only if their own
systematic uncertainties were at the 1\% level, and we have already
noted larger broad band discrepancies between the \heg\ and \meg.

\begin{deluxetable*}{cccccccccc}  
\setlength{\tabcolsep}{0.03in} 
\tabletypesize{\footnotesize}    
\tablewidth{0pt} 
\tablecaption{{\tt kerrbb+comptt+gaussian} Fits to Selected RXTE Data \label{tab:rxtekrra}} 
\tablehead{ \colhead{ID}                    
           & \colhead{$a$}                   
           & \colhead{${\rm \dot M}$}      
           & \colhead{${\rm h_d}$}          
           & \colhead{${\rm D_{BH}}$}       
           & \colhead{${\rm A_{comptt}}$}   
           & \colhead{$\tau_{\rm p}$}      
           & \colhead{${\rm A_{line}}$}     
           & \colhead{$\sigma_{\rm line}$} 
           & \colhead{$\chi^2/$dof}         
           \\                               
           & & ($10^{17}~{\rm g~s^{-1}}$)    
           & & (kpc)                          
           & $(10^{-5})$                      
           & & $(10^{-4})$                    
           & (keV)                            
           }                                  
\startdata  
40044-01-02-01 % ObsID 
           & \errtwo{ 0.99026}{ 0.00473}{ 0.00000} % BH Spin  
           & \errtwo{ 0.749}{ 0.171}{ 0.113} % Mass accretion rate  
           & \errtwo{ 0.929}{ 0.045}{ 0.043} % Hardening factor 
           & {\sl 10} % BH distance 
           & \errtwo{263.7}{ 7.5}{ 8.5} % Comptt norm  
           & \errtwo{ 0.19}{ 0.02}{ 0.01} % taup  
           & \errtwo{  5.4}{  4.6}{  3.0} % Line norm  
           & \errtwo{  0.5}{  0.5}{  0.5} % Line sigma  
           & 18.0/27 % Chi^2/dof  
\\ 
\tabspace  
           &  \errtwo{ 0.99409}{ 0.00581}{ 0.99409} % BH Spin  
           & \errtwo{ 1.541}{ 4.817}{ 0.761} % Mass accretion rate  
           & {\sl 1.7} % Hardening factor 
           & \errtwo{15.1}{56.1}{ 2.8} % Distance 
           & \errtwo{264.0}{ 7.5}{10.4} % Comptt norm  
           & \errtwo{ 0.19}{ 0.02}{ 0.01} % taup  
           & \errtwo{  5.5}{  5.2}{  3.1} % Line norm  
           & \errtwo{  0.5}{  0.5}{  0.5} % Line sigma  
           & 18.0/27 % Chi^2/dof  
\\ 
\tabspace  
           &  \errtwo{ 0.59987}{ 0.08325}{ 0.08022} % BH Spin  
           & \errtwo{ 3.556}{ 0.471}{ 0.452} % Mass accretion rate  
           & {\sl 1.7} % Hardening factor 
           & {\sl 10}  % Distance 
           & \errtwo{254.2}{ 7.9}{16.5} % Comptt norm  
           & \errtwo{ 0.20}{ 0.01}{ 0.02} % taup  
           & \errtwo{  4.3}{  3.9}{  2.2} % Line norm  
           & \errtwo{  0.3}{  0.4}{  0.3} % Line sigma  
           & 19.0/28 % Chi^2/dof  
\\ 
\tabspace  
70054-01-11-00 % ObsID 
           & \errtwo{ 0.99896}{ 0.00002}{ 0.00009} % BH Spin  
           & \errtwo{ 0.349}{ 0.001}{ 0.001} % Mass accretion rate  
           & \errtwo{ 1.102}{ 0.002}{ 0.003} % Hardening factor 
           & {\sl 10} % BH distance 
           & \errtwo{ 0.0}{ 0.2}{ 0.0} % Comptt norm  
           & \errtwo{ 0.01}{ 4.99}{ 0.00} % taup  
           & \errtwo{  3.9}{  0.4}{  0.6} % Line norm  
           & \errtwo{  0.8}{  0.2}{  0.2} % Line sigma  
           & 16.7/17 % Chi^2/dof  
\\ 
\tabspace  
           &  \errtwo{ 0.99986}{ 0.00001}{ 0.00001} % BH Spin  
           & \errtwo{ 1.161}{ 0.002}{ 0.004} % Mass accretion rate  
           & {\sl 1.7} % Hardening factor 
           & \errtwo{19.5}{ 0.1}{ 0.0} % Distance 
           & \errtwo{ 4.6}{ 0.2}{ 0.4} % Comptt norm  
           & \errtwo{ 0.01}{ 0.01}{ 0.00} % taup  
           & \errtwo{  7.1}{  0.4}{  0.7} % Line norm  
           & \errtwo{  1.0}{  0.0}{  0.1} % Line sigma  
           & 25.7/17 % Chi^2/dof  
\\ 
\tabspace  
           &  \errtwo{ 0.82851}{ 0.00334}{ 0.01372} % BH Spin  
           & \errtwo{ 1.414}{ 0.035}{ 0.019} % Mass accretion rate  
           & {\sl 1.7} % Hardening factor 
           & {\sl 10}  % Distance 
           & \errtwo{ 0.0}{0.4}{ 0.0} % Comptt norm  
           & \errtwo{ 0.01}{ 4.99}{ 0.00} % taup  
           & \errtwo{  4.5}{  1.7}{  1.0} % Line norm  
           & \errtwo{  0.7}{  0.2}{  0.2} % Line sigma  
           & 46.5/18 % Chi^2/dof  
\\ 
\tabspace  
90123-01-03 % ObsID 
           & \errtwo{ 0.99969}{ 0.00021}{ 0.01854} % BH Spin  
           & \errtwo{ 0.576}{ 0.467}{ 0.017} % Mass accretion rate  
           & \errtwo{ 1.087}{ 0.017}{ 0.014} % Hardening factor 
           & {\sl 10} % BH distance 
           & \errtwo{ 5.6}{ 4.6}{ 5.1} % Comptt norm  
           & \errtwo{ 0.02}{ 2.50}{ 0.01} % taup  
           & \errtwo{  4.7}{  1.9}{  1.5} % Line norm  
           & \errtwo{  0.5}{  0.3}{  0.4} % Line sigma  
           & 25.0/25 % Chi^2/dof  
\\ 
\tabspace  
           &  \errtwo{ 0.99974}{ 0.00012}{ 0.00014} % BH Spin  
           & \errtwo{ 2.676}{ 0.154}{ 0.016} % Mass accretion rate  
           & {\sl 1.7} % Hardening factor 
           & \errtwo{21.7}{ 0.4}{ 0.1} % Distance 
           & \errtwo{ 6.0}{ 0.7}{ 3.0} % Comptt norm  
           & \errtwo{ 0.01}{ 4.99}{ 0.00} % taup  
           & \errtwo{  4.6}{  0.8}{  0.8} % Line norm  
           & \errtwo{  0.5}{  0.2}{  0.2} % Line sigma  
           & 24.9/25 % Chi^2/dof  
\\ 
\tabspace  
           &  \errtwo{ 0.84529}{ 0.00236}{ 0.00463} % BH Spin  
           & \errtwo{ 2.412}{ 0.033}{ 0.029} % Mass accretion rate  
           & {\sl 1.7} % Hardening factor 
           & {\sl 10}  % Distance 
           & \errtwo{ 0.0}{2.1}{ 0.0} % Comptt norm  
           & \errtwo{ 0.37}{ 4.63}{ 0.36} % taup  
           & \errtwo{  9.0}{  1.5}{  1.5} % Line norm  
           & \errtwo{  0.7}{  0.1}{  0.1} % Line sigma  
           & 48.0/26 % Chi^2/dof  
\\ 
\tabspace  
90063-01-01 % ObsID 
           & \errtwo{ 0.99852}{ 0.00019}{ 0.00312} % BH Spin  
           & \errtwo{ 0.655}{ 0.042}{ 0.021} % Mass accretion rate  
           & \errtwo{ 1.113}{ 0.021}{ 0.011} % Hardening factor 
           & {\sl 10} % BH distance 
           & \errtwo{ 6.2}{ 0.8}{ 0.1} % Comptt norm  
           & \errtwo{ 0.01}{ 0.68}{ 0.00} % taup  
           & \errtwo{  5.1}{  1.6}{  1.7} % Line norm  
           & \errtwo{  0.6}{  0.2}{  0.3} % Line sigma  
           & 18.8/25 % Chi^2/dof  
\\ 
\tabspace  
           &  \errtwo{ 0.99795}{ 0.00050}{ 0.00268} % BH Spin  
           & \errtwo{ 3.708}{ 0.474}{ 0.472} % Mass accretion rate  
           & {\sl 1.7} % Hardening factor 
           & \errtwo{23.3}{ 0.2}{ 1.5} % Distance 
           & \errtwo{ 5.2}{ 1.7}{ 3.8} % Comptt norm  
           & \errtwo{ 0.03}{ 4.97}{ 0.02} % taup  
           & \errtwo{  5.1}{  1.8}{  1.7} % Line norm  
           & \errtwo{  0.6}{  0.2}{  0.3} % Line sigma  
           & 18.8/25 % Chi^2/dof  
\\ 
\tabspace  
           &  \errtwo{ 0.84516}{ 0.00203}{ 0.00401} % BH Spin  
           & \errtwo{ 2.415}{ 0.035}{ 0.026} % Mass accretion rate  
           & {\sl 1.7} % Hardening factor 
           & {\sl 10}  % Distance 
           & \errtwo{ 0.0}{2.3}{ 0.0} % Comptt norm  
           & \errtwo{ 2.50}{ 2.50}{ 2.49} % taup  
           & \errtwo{  8.7}{  1.6}{  1.3} % Line norm  
           & \errtwo{  0.7}{  0.1}{  0.1} % Line sigma  
           & 39.0/26 % Chi^2/dof  
\\ 
\tabspace  
\enddata 
\tablecomments{One fit used a fixed mass of 3\,M$_\odot$, a 
fixed distance of 10\,kpc, and a variable spectral hardening factor, 
another set used a fixed mass of 16\,M$_\odot$, a fixed 
spectral hardening factor of 1.7, and a variable distance, and the third set
a fixed mass of 3\,M$_\odot$, a fixed distance of 10\,kpc, and a fixed 
spectral hardening factor of 1.7.  Line normalization, ${\rm A_{line}}$, 
is total photons/cm$^2$/s in the line. 
0.5\% systematic errors were added in quadrature to the data.
Error bars are 90\% confidence level.} 
\end{deluxetable*}

%%%%%% Fig. 9 %%%%%

\begin{figure*}
\epsscale{1}
\plotone{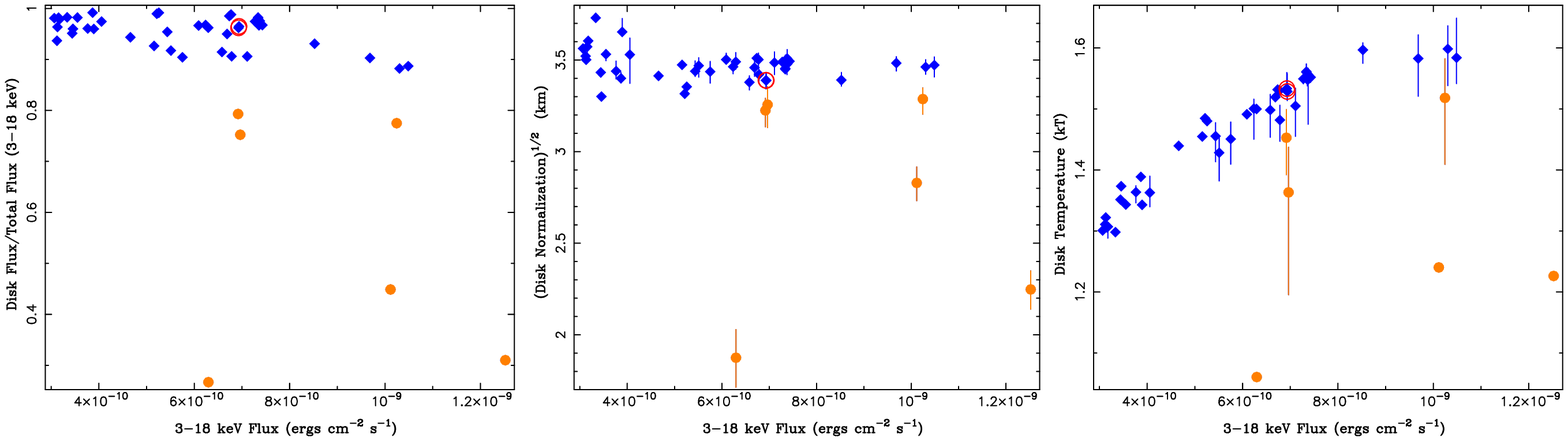}
\caption{Left: The fraction of disk flux to total flux, in the
  3-18\,keV energy band, for {\tt diskbb+powerlaw} fits to \fu.
  Middle: Square root of disk normalization vs. \pca\ flux for {\tt
    diskbb+powerlaw} fits to \rxte\ observations of \fu.  Right: Disk
  temperature vs. \pca\ for {\tt diskbb+comptt} fits to \fu.
\label{fig:rxte_flux}}
\end{figure*}  

For the disk plus power-law fits, we constrained the power-law photon
index, $\Gamma$, to lie between 1--3, but let the power-law
normalization freely vary.  These fits were fairly successful overall
(reduced $\chi^2$ ranged from 0.5--2, and averaged 1).  Parameter
trends with observed flux are shown in Fig.~\ref{fig:rxte_flux}, and
show several features of interest.  First, the vast majority of the
observations are consistent with having a nearly constant disk
normalization, which one could take as a proxy for disk radius.  At
the low end of 3--18\,keV flux, the disk normalization/radius turns
upward.  Phenomenologically, such observed increases in disk radius
have been associated with low luminosity transitions to the hard state
\citep[e.g., see the similar behavior of LMC~X-3;][]{wilms:01a}, which
we expect to occur near 3\% of the Eddington luminosity
\citep[][]{maccarone:03a}.  It is therefore possible that our lowest
luminosity observations are near such a transition at 3\%~${\rm
  L_{Edd}}$.  (On the other hand, \citealt{saito:06a} note a similar
increase in fitted disk radius for GRO J1655$-$40, without any
associated transition, or near transition, to a low/hard state.)

Second, we see that as the 3--18\,keV flux increases, the fraction of
the flux contributed by the power-law slightly increases.  Half a
dozen points, however, show a much more significant contribution by
the power-law (Fig.~\ref{fig:rxte_flux}).  At the same time, those
observations show a dramatic decrease in the disk normalization. Each
of those observations is associated with peaks in the ASM lightcurve,
and several are associated with high variability, including one
instance of ``rapid state transitions'' (see \S\ref{sec:rxte_vary}).
We identify those observations with what \citet{remillard:06a} refer
to as the ``steep power-law'' state (or, the ``very high state'';
\citealt{miyamoto:93a}).

%%%%% Fig. 10 %%%%

\begin{figure*}
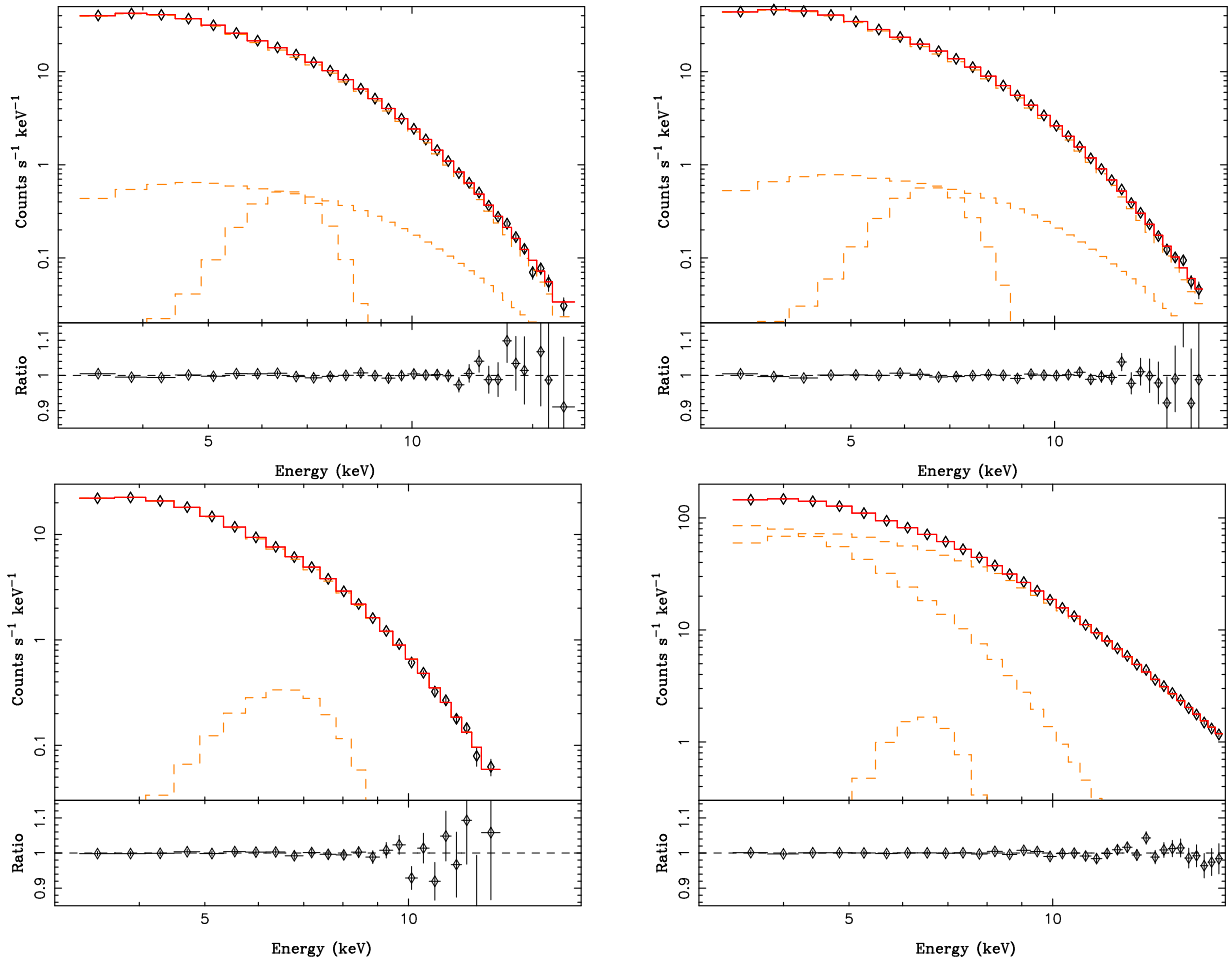

\epsscale{1}
\plottwo{f10a_col.ps}{f10b_col.ps}
\plottwo{f10c_col.ps}{f10d_col.ps}
\caption{Count rate spectra for the \rxte\ observations simultaneous
  with \chandra\ (top left) and \xmm\ (top right), and for the
  softest/faintest (bottom left) and hardest/brightest (bottom right)
  of the \rxte\ observations, all fit with the {\tt
    kerrbb+comptt+gaussian} model (3\,$\msun$, 10\,kpc, and freely
  varying color correction fraction). Dashed lines show the individual
  model components folded through the detector response.
\label{fig:kerr_fits}}
\end{figure*}  

Even though the disk normalization drops and the power-law flux
increases for the disk plus power-law fits, the fitted disk
temperature does not show any significant deviations from a trend of
3--18\,keV flux $\propto (kT)^5$ (not shown in
Fig.~\ref{fig:rxte_flux}).  For a disk spectrum with constant disk
radius, we expect the bolometric flux to scale $\propto (kT)^4$, and,
for these temperatures, to scale approximately as $(kT)^5$ in the
3--18\,keV bandpass.  This same relation between disk temperature and
3--18\,keV flux holds if we instead fit the spectra with a disk plus
Comptonization spectrum, as shown in Fig.~\ref{fig:rxte_flux}.

For these latter fits, we model the spectrum in the same manner as in
\S\ref{sec:broad}: we use a {\tt diskbb} {\tt +} {\tt comptt} model
and tie the seed photon temperature to the disk peak temperature,
freeze the corona temperature to 50\,keV, but let the coronal spectrum
normalization and optical depth (constrained by $0.01 < \tau_{\rm p} <
5$) be fit parameters.  Results for several of these fits are
presented in Table~\ref{tab:rxtedbb}.  The majority of the
observations again show a nearly constant disk radius, and a
3--18\,keV flux that scales approximately $\propto (kT)^5$.  The
Comptonization models, however, now show the half dozen `steep
power-law state' observations as being associated with dramatic drops
in the disk/seed photon temperatures (Fig.~\ref{fig:rxte_flux}).

We now consider fits where the unphysical {\tt diskbb} component is
replaced with the {\tt kerrbb} model.  As for the fits to the
\chandra\ data, we fix the disk inclination to 75$^\circ$, and set the
source distance to 10\,kpc and the black hole mass to 3\,$\msun$.  The
latter choices are motivated by our faintest \rxte\ observation for
which we measure a 3--18\,keV absorbed flux of $3.1\times10^{-10}~{\rm
  ergs~sec^{-1}~cm^{-2}}$.  Taking the {\tt diskbb} model at face
value, with an inclination of $75^\circ$, this result translates to an
unabsorbed bolometric luminosity of 3\%\,${\rm L_{Edd}}$ for this
presumed mass and distance.  The black hole is unlikely to be less
massive than 3\,$\msun$, and low black hole masses lead to higher disk
temperatures.  Thus, we see these parameter choices as
\emph{minimizing} the need for high spin parameters in the {\tt
  kerrbb} fits, with 10\,kpc then being a \emph{lower limit} to the
source distance.

The remaining {\tt kerrbb} fit parameters are the accretion rate,
$\dot {\rm M}$, the black hole spin, $a^*$, and the disk atmosphere
color-correction factor, ${\rm h_d}$.  The {\tt kerrbb} model does not
fit a disk temperature per se; therefore, we set the input seed photon
temperature of the {\tt comptt} model equal to the {\tt diskbb} peak
temperature from our previous fits \citep[see][]{davis:06a}.  As
before, we also freeze the coronal temperature to 50\,keV, and
constrain the optical depth to lie between 0.01--5.  Selected fit
results are presented in Table~\ref{tab:rxtekrra} and
Fig.~\ref{fig:kerr_fits}.

\begin{figure}
\epsscale{1}
\plotone{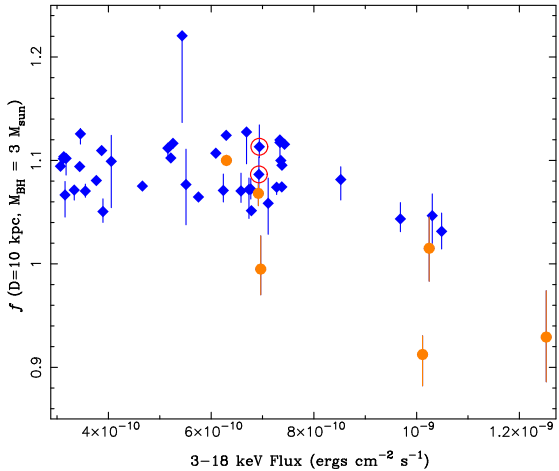}
\caption{Spectral hardening factor vs. \pca\ flux for {\tt
    kerrbb+comptt} fits to \fu.  We assume a black hole mass of 3
  ${\rm M}_\odot$, a distance of 10\,kpc, and an inclination of 75$^\circ$.
\label{fig:kerrbb_hd}}
\end{figure}

Even with choosing a very low black hole mass, and thus naturally
favoring higher disk temperatures, the {\tt kerrbb} models uniformly
prefer to fit $a^*\approx 1$.  On the other hand, the disk atmosphere
spectral hardening remains low, with ${\rm h_d} \approx 1.1$. There is
a slight trend for ${\rm h_d}$ to decrease with increasing flux, as
shown in Fig.~\ref{fig:kerrbb_hd}.  For some of the observations where
the disk contribution to the total flux dramatically decreased, ${\rm
h_d}$ drops slightly lower still.  This is not surprising as the
inclusion of a coronal component can qualitatively replace the need
for hardening from the disk atmosphere.

The {\tt kerrbb} model has a number of parameter degeneracies which we
can exploit to search for fits with the more usual color-correction
factor of ${\rm h_d} = 1.7$.  Specifically, the apparent temperature
increases $\propto {\rm h_d}$, and $\propto (\dot {\rm M}/{\rm
  M}^2)^{1/4}$.  If we increase ${\rm h_d}$, then in order to retain
roughly the same spectral shape we must decrease $({\rm \dot M}/{\rm
  M}^2)^{1/4}$. Solely decreasing ${\rm \dot M}$, however, makes our
faintest observation less than 3\%\,${\rm L_{Edd}}$. Additionally, the
source distance would need to be reduced $\propto {\rm \dot M}^{1/2}$
in order to retain the same observed flux. We would have expected to
detect a transition to a hard spectral state if ${\rm \dot M}$ were
lower than we have assumed.  In order to keep our faintest observation
at $\approx 3\%$\,${\rm L_{Edd}}$, we must instead scale ${\rm M}
\propto {\rm h_d}^4$, $\dot {\rm M} \propto {\rm h_d}^4$, and the
source distance as $\propto {\rm h_d}^{2}$.  A consistent picture may
be obtained with ${\rm h_d} = 1.7$, if $M \approx 16$\,$\msun$ and the
distance to the source is $\approx 22$\,kpc.

%%%%% Fig. 11 %%%%

\begin{figure}
\epsscale{1}
\plotone{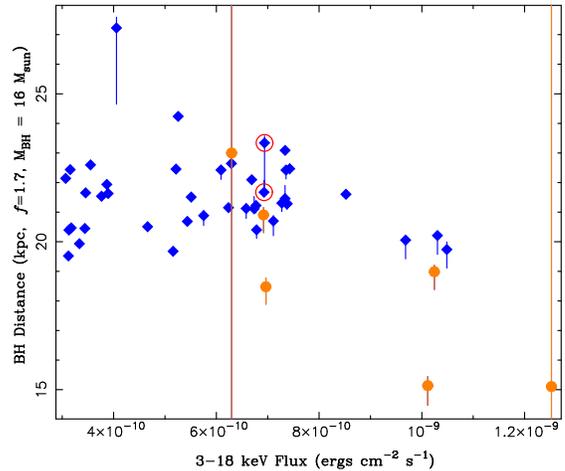}
\caption{Fitted black hole distance for {\tt kerrbb+comptt} fits to
  \fu, assuming a spectral hardening factor of ${\rm h_d} = 1.7$, a
  black hole mass of 16 ${\rm M}_\odot$, and an inclination of 75$^\circ$.
\label{fig:kerrbb_d}}
\end{figure}

We searched for a set of {\tt kerrbb+comptt} fits where we froze the
disk color-correction factor at ${\rm h_d}=1.7$, the black hole mass
at 16\,$\msun$, but let the disk accretion rate, $\dot {\rm M}$, spin
parameter, $a^*$, and source distance all be free parameters.
Clearly, the source distance cannot truly be physically changing in a
perceptible way; however, the degree to which we obtain a uniform set
of fitted distances might provide a self-consistency check on the use
of the {\tt kerrbb} model.  Selected results from these fits are
presented in Table~\ref{tab:rxtekrra}, and the fitted distances
vs. 3--18\,keV flux are presented in Fig.~\ref{fig:kerrbb_d}. In
general, these fits work every bit as well as the 3\,$\msun$, 10\,kpc
fits.  The lower flux observations cluster about a fitted distance of
$\approx 22$\,kpc.  There is a trend, however, for the fitted
distances to decrease with higher fluxes, and for several of the
`steep power-law state' observations to fit distances as low as
15\,kpc.

We have considered another class of fit degeneracy: disk atmosphere
hardening factor and spin.  Here we again freeze the black hole mass
to 3\,$\msun$ and the distance to 10\,kpc, but freeze the hardening
factor to ${\rm h_d} = 1.7$.  As more of the peak in the apparent
temperature is being attributed to the color-correction factor, the
disk spin must be decreased, thereby decreasing the accretion
efficiency and increasing the radius of the emitting area near the
peak temperatures.  To maintain the same observed flux, the accretion
rate must be increased above that for the models with lower ${\rm
  h_d}$.  Formally, these are significantly worse fits than our
previous ones, with a mean $\Delta \chi^2 = 22$.  I.e., an increased
accretion rate and spectral hardening factor are not acting as simple
proxies for high spin in these models. (In terms of fractional
residuals, however, the fits are somewhat reasonable.)

\begin{figure}
\epsscale{1}
\plotone{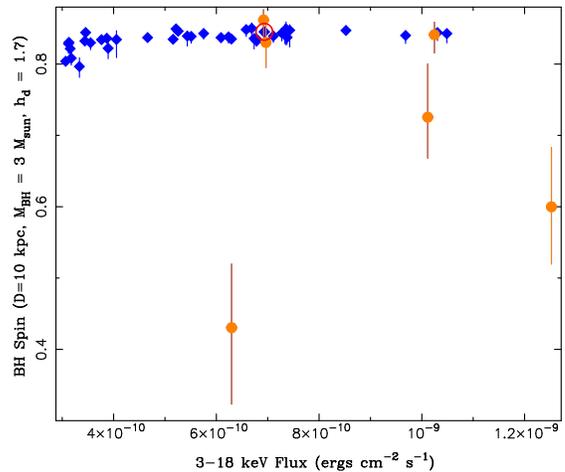}
\caption{Fitted black hole spin for {\tt kerrbb+comptt} fits to \fu,
  assuming a spectral hardening factor of ${\rm h_d} = 1.7$, a black
  hole mass of 3\,${\rm M}_\odot$, a distance of 10\,kpc, and an inclination
  of 75$^\circ$.
\label{fig:kerrbb_a}}
\end{figure}

As shown in Fig.~\ref{fig:kerrbb_a}, these fits yield a relatively
uniform spin of $a^* \approx 0.84$, with a slight trend for fitted
spin to increase with increasing flux.  Notable exceptions to this
behavior are seen, however, as some of the `steep power-law state'
observations fit much reduced black hole spins.  Again, this is a
contribution from a Comptonization component, which is strong for
these observations, replacing some of the need for spectral hardening
due to rapid spin.

%%%%%% Fig. 12 %%%%%%

Note that we did search for a set of disk atmosphere model fits where
we kept the source mass low ($3\,{\rm M}_\odot$) and moved the distance
further out (17\,kpc) while fitting higher accretion rates, to
intrinsically increase the disk temperatures and raise the implied
fractional Eddington luminosities by a factor of $\approx 3$.  These
models \emph{failed completely, with best fit $\chi^2$ values
  increasing by factors of almost 3 over high-spin models}.  We again
find that high-spin is \emph{not} simply degenerate with increased
accretion rate in the models.  Essentially this is because high-spin
adds an inner disk region with high temperature and high flux
(approximately tripling the flux when going from non-spinning to Kerr
solutions), which leads to a strong, broad component in the high
energy spectrum. High accretion, low spin models, fail to reproduce
such broad, high energy \pca\ spectra.

\subsection{Variability}\label{sec:rxte_vary}

The study of the rapid variability properties of \fu\ was performed
using high time resolution data from the \pca. We created power
spectra from the entire \pca\ energy band ($\approx$2--60 keV),
covering the 2$^{-7}$--2048 Hz frequency range. Observations with
high disk fractions ($>$80\%, as defined by Fig.~\ref{fig:rxte_flux})
typically showed weak or no significant variability. Combining these
observations resulted in a power spectrum that could be fitted
reasonably well by a single power-law component
($\chi^2$/dof=101.0/79). This fit improved significantly
($\chi^2$/dof=73.1/74) by adding two Lorentzians, one (with Q-value
fixed at 0 Hz) for a weak band-limited noise component around 2\,Hz
and the other for a QPO at 25\,Hz.  The resulting parameters from the
fit are listed in Table~\ref{tab:pds_fit}. Note, that although the
improvement in the fit is statistically significant, the two added
components are only marginally significant themselves
($\approx3\sigma$).

\begin{figure}
\epsscale{1}
\plotone{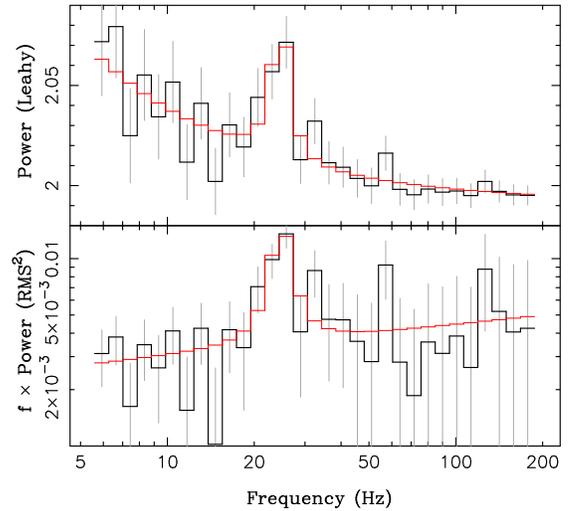}
\caption{An approximately 25\,Hz quasi-periodic oscillation seen in
  the 4.5-27.5\,keV \fu\ lightcurve from ObsID 50128-01-09-00, the
  observation with the smallest disk contribution to the observed
  3-18\,keV flux.  The top panel shows the power spectrum with the
  Poisson noise level unsubtracted and normalized to $\approx 2$.  The
  bottom panel shows Fourier frequency times the power spectrum, now
  with the Poisson noise level subtracted and the PSD normalized to
  (root mean square)$^2$/Hz.
\label{fig:qpo}}
\end{figure}

Observations with lower disk fractions showed an increase in
variability, with maximum of 12.1$\pm$0.6\% root mean square (rms)
variability (0.01--100 Hz) in the observation with the lowest disk
fraction (ObsID 50128-01-09-00; disk fraction of $\approx$30\%; see
Fig.~\ref{fig:qpo}). Combining the power spectra of all observations
with disk fractions lower than 80\% and fitting with the same model as
above, we find that all variability components increased in strength
(see Table~\ref{tab:pds_fit}).

%%%%%%%% Fig. 13 %%%%%%%

The power spectral properties of the high- and low-disk fraction
observations are consistent with the soft state and the soft end of
the transition between the hard and soft states,
respectively. Observation of such variability strengthens the argument
that \fu\ is indeed a black hole, and not a neutron star.  The
frequencies of the QPOs are similar to that seen in a few other black
hole systems when they were in or close to their soft state, e.g. XTE
J1550$-$564 \citep{homan:01a} and GRO J1655$-$40 \citep{sobczak:00a}.
The comparable frequencies observed in these latter two sources have
ranged from 15--22\,Hz, and futhermore the spectra of these sources
have been fit with disk models that imply black hole spins in the
range of $a^* \approx 0.1$--0.8 \citep{davis:06a,shafee:06a}.  Thus,
if such variability features are related to black hole spin, a
consistent picture arises between their spectral and variability
results.

In two of the low-disk fraction observations (ObsIDs 40044-01-02-01
and 70054-01-04-00) we observed fast changes in the count rate, which
resemble the dips/flip-flops observed in other black hole X-ray
binaries during transition from the hard to the soft state
\citep[e.g.,][]{miyamoto:91a}.  In \fu, these changes in the count
rate were accompanied by moderate changes in the spectral variability
properties, but count rates were too low to classify the power spectra
in the high and low count rate phases independently.

\section{Discussion}\label{sec:discuss}

We have presented observations of the black hole candidate
\fu\ performed with the \chandra, \xmm, and \rxte\ X-ray
observatories.  All of these observations point toward a relatively
simple and soft spectrum, indicative of a classic BHC disk-dominated
soft state.  The \chandra\ and \xmm\ spectra, especially at high
resolution, indicate a remarkably unadulterated disk spectrum.  That
is, the absorption of the spectrum is very low at only ${\rm N_H} =
1$--$2\times 10^{21}\,{\rm cm}^{-2}$, and with the possible exception
of an unidentified line at 21.8\,\AA, there is very little evidence of
spectral complexity intrinsic to the source.  From that vantage point,
\fu\ may be the cleanest disk spectrum with which to study modern disk
atmosphere models.

\begin{deluxetable}{lcc}
\setlength{\tabcolsep}{0.03in} 
\tabletypesize{\scriptsize}    
\tablewidth{0pt} 
\tablecaption{Power spectra fit parameters\label{tab:pds_fit}}
\tablehead{  \colhead{Parameter\tablenotemark{a}} 
           & \colhead{High Disk Frac.} 
           & \colhead{Low Disk Frac.} }
\startdata
PL rms (\%)           & 1.05$\pm$0.09          & 2.6$^{+0.7}_{-0.5}$ \\
\tabspace
PL index              & 1.57$\pm$0.15          & 0.92$\pm$0.12       \\
\tabspace
BLN rms (\%)          & 1.2$\pm$0.2            & 4.1$^{+1.1}_{-0.7}$ \\
\tabspace
BLN $\nu_{max}$ (Hz)  & 1.9$^{+1.5}_{-0.8}$    & 13$^{+9}_{-5}$      \\
\tabspace
QPO rms (\%)          & 2.0$\pm$0.3            & 3.4$\pm$0.9         \\
\tabspace
QPO Q-value           & 2.10$^{+1.24}_{-0.76}$ & 2.5$^{+3.6}_{-1.1}$ \\
\tabspace
QPO $\nu_{max}$ (Hz)  & 25.0$\pm$2.2           & 29.5$\pm$2.1        \\
total rms\tablenotemark{b}         & 2.7$\pm$0.4            & 6.1$\pm$0.4         \\
$\chi^2$/dof        & 73.1/74                & 65.9/80             \\
\enddata

\tablenotetext{a}{PL = power-law; BLN = band-limited noise; QPO = quasi-periodic oscillation}
\tablenotetext{b}{0.01--100 Hz}
\end{deluxetable}

Perhaps the one indication of additional, unmodeled spectral
complexity is the fact that the broadband fits and the more localized,
high resolution edge and line fits yield (low) neutral columns that
differ from each other by a factor of two.  This, along with the
21.8\,\AA\, absorption feature might indicate the presence of other
weak, unmodeled spectral components at very soft X-ray energies.  Even
if this were the case, this is to be compared to other BHC to which
such atmosphere models have been applied, e.g., GRO J1655$-$40
\citep{shafee:06a}, where there is both a larger neutral column (${\rm
  N_H} = 7 \times 10^{21}\,{\rm cm^{-2}}$), a complex Fe K$\alpha$
region (modeled with both emission and smeared edges by
\citealt{shafee:06a}) and high resolution observations of
intermittently present, complex X-ray spectral features that are local
to the system \citep[i.e.,][]{miller:06a,miller:08a}. In comparison,
the \fu\ spectrum, even with its uncertainties, is much simpler.

The observations presented here strengthen the case that \fu\ is a
black hole system.  The spectrum closely matches a classic, soft state
disk spectrum (well-modeled by {\tt diskbb}), with a contribution from
a harder, non-disk component that generally increases as the flux
increases.  Occasionally (a half dozen observations), the contribution
from this hard component increases, the X-ray variability increases, a
possible QPO is observed, and we even observe lightcurve `dipping' and
`flip-flop' behavior \citep{miyamoto:91a,miyamoto:93a}.  All of this
is behavior familiar from studies of other BHC \citep[][and references
  therein]{homan:01a}.  Unfortunately, observation of these behaviors
provides little strong input to estimates of the system parameters,
i.e., mass, distance, and inclination.  `Flip-flop' behavior has often
been associated with luminosities of a few tens of percent
\citep{miyamoto:91a,miyamoto:93a,homan:01a}, which is higher than we
have presumed for the fits described here. The threshold luminosity
for such behavior, however, is not firm, and we further note that high
accretion rate/large distance fits to the \pca\ spectra failed
(\S\ref{sec:rxte_spectra}).  The fact that we do not see a transition
to a BHC hard state--- although we may see some indications from
increases in the fitted disk radius at low luminosity--- strongly
suggest that the mass is $>3\,\msun$ and the distance is $>10$\,kpc.

Despite the fact that we do not have a firm estimate of mass and
distance for \fu, the relativistic disk models still strongly
statistically prefer rapid spin solutions.  Essentially, this is
because the {\tt diskbb} models fit very high temperatures -- as high
as 1.7\,keV -- with fairly low normalizations.  We saw, however, that
for the \chandra\ observation, or for the \rxte\ observations where
the power-law/Comptonization component strengthens, the need for
spectral hardening of the disk, either via a color-correction factor
(${\rm h_d}$) or rapid spin, was greatly reduced.  Phenomenologically,
a coronal component can mimic the effects of rapid spin.  This
naturally leads to the question of whether or not one is sure that any
residual coronal component completely vanishes at low flux.  Does one
ever observe a ``bare'' disk spectrum?

It is a rather remarkable fact that the \chandra\ observation, and the
low flux \rxte\ observations, are mostly described via three
parameters: absorption, {\tt diskbb} temperature, and {\tt diskbb}
normalization.  Even more remarkably, changes in the spectrum, both in
terms of amplitude and overall shape, are mainly driven by variations
of only \emph{one} parameter: {\tt diskbb} temperature.  The
relativistic disk models, on the other hand, must ascribe essentially
this one parameter to a combination of four parameters: $a^*$, disk
inclination (which affects the appearance of relativistic features),
$({\rm \dot M/M}^2)$, and ${\rm h_d}$.  In principle, any of the
latter three can vary among observations.  (Disk inclination can vary
due to warping; see \citealt{pringle:96a}.)  Unfortunately, there is
no truly unique, sharp feature in the high resolution spectrum that
completely breaks degeneracies among these parameters. The overall
magnitude of the {\tt diskbb} temperature and shape of the
\fu\ spectra, however, seem best reproduced with the most rapid spin
models; simply increasing accretion rates or spectral hardening
factors is insufficient to model the spectra fully.

We do not see the essential question as being whether or not \fu\ is
rapidly spinning.  Instead, we view the more observationally motivated
and perhaps more fundamental question as being: why is the
characteristic disk temperature of \fu\ so high?  Rapid spin is one
hypothesis.  Another would be that there is indeed a residual, low
temperature corona, even at low flux.  The possibly high inclination
of this source ($\approx 75^\circ$ to be consistent with optical
lightcurve variations, while being consistent with the lack of X-ray
eclipses) would mean that we are viewing the disk through a larger
scattering depth than would be usual for most BHC in the soft state.
We also note that there exists very little, self-consistent
theoretical modeling of low temperature Comptonizing coronae with high
temperature (1--2\,keV) seed photon temperatures.  Most energetically
balanced, self-consistent models \citep[e.g.,][]{dove:97a} have focused
on high coronal temperatures (50--200\,keV) with low seed photon
temperatures (100--300\,eV).

If one is to take the rapid spin hypothesis for the high {\tt diskbb}
temperature at face value, then these observations offer something of
a prediction.  Statistically, they do prefer rapid spin, and based
upon observations of other BHC we would have expected to observe a
hard state if the faintest observations were $\aproxlt 3\%\,{\rm
L_{Edd}}$.  If the theoretically preferred value of the spectral
hardening factor is indeed ${\rm h_d}=1.7$, future measurements should
find \fu\ to be an $\approx 16\,\msun$ black hole at $\approx
22$\,kpc.  Whether or not this turns out to be the case, given the
nature of \fu\ as perhaps the simplest, cleanest example of a BHC soft
state, observations to independently determine this system's
parameters (mass and distance) are urgently needed.

\acknowledgments It is a pleasure to acknowledge useful conversations
with Manfred Hanke, David Huenemoerder, John Houck, and John Davis.
This work has been supported by NASA grant SV3-73016 and DLR grant
50OR0701.

\appendix

In this appendix, we describe how we model the mild pile up that is
present in the \meg\ and, to lesser degree, the \heg\ data. Our model
follows the discussions of \citet{davis:01a} and \citet{davis:03a},
although we do not attempt to arrive at an {\sl ab initio} description
of pile up. Instead, we present a simple, phenomenological description
where the absolute amplitude of pile up is left as a fit parameter
(albeit one that can be determined roughly with empirical knowledge
and then frozen at a fixed value, or fitted with a limited range of
``acceptable'' values).  In contrast to spectral pile up solely with
the CCD detector, where piled events can reappear in the spectrum at
higher implied energies, gratings pile up (at least in \fst\ order
spectra) can be thought of as a straight loss of events. Whether the
piled events `migrate' to bad grades, or are read as a single event of
higher energy and are thus removed from the `order sorting window', it
is still a simple loss term \citep{davis:03a}.

The degree of loss due to pile up scales with $C_i$, the number of
expected incident events in a given detector region per detector frame
integration time.  The detected number of events is then reduced by a
factor $\exp({\cal A} C_i)$, for a suitably chosen region and where
${\cal A}$ is a `fudge factor' of order unity.  The proper detector
region to consider is approximately 3 pixels in the `cross dispersion'
direction, by a 3--5 pixel length along the dispersion direction of
the gratings arm being analyzed. (Events in a single frame that land
in adjacent pixels will always be piled, while events that are removed
from one another by four pixels will only be piled if their charge
clouds extend toward one another; see \citealt{davis:03a}.)  Given
that the peak effective area of the \meg\ is approximately twice that
of the \heg, and the fact that detector pixels cover twice the
wavelength range in the \meg, we expect the pile up to be
approximately 4 times larger in the \meg.

It is important to note that one needs to consider \emph{all} events
in a given detector region, i.e., all spectral orders, background
events, contributions from the wings of the \zth\ order point spread
function, etc., whether or not these events are otherwise rejected by
order sorting.  To partly account for this necessity, the pile up
model described here uses information from the \scnd\ and \trd\ order
ancillary response functions (\arfs).  Specifically, we create a
`convolution model' where, given an input spectrum, we define
$C_i(\lambda)$ as the first order \arf\ multiplied by the unpiled
model spectrum (yielding counts per detector bin).  We further
increase $C_i(\lambda)$ by multiplying the unpiled model by the \scnd\
order \arf, then shifting the \scnd\ order wavelengths by a factor of
two, and then rebinning and adding this contribution to
$C_i(\lambda)$.  A similar procedure is used for the \trd\ order
contribution.  We then use the input pile up fraction fit parameter,
$p_f$, to normalize the exponential reduction of the spectrum.
Specifically, we multiply the counts per bin predicted for the unpiled
model by $\exp[\log(1-p_f)[C_i(\lambda)/\max(C_i)]]$.

There are two important subtleties to address.  First, in certain
regions of the spectrum, specifically those wavelength regions that
are dithered over chip gaps or detector bad pixels, the
counts are lower not due to a low intrinsic flux, but rather due to a
fractionally reduced exposure.  The {\tt FRACEXPO} column found in the
\arf\ FITS files for \chandra\ gratings data provides this
information; therefore, we can approximately correct for this
effect. Second, for \chandra\ gratings spectra, a portion of the
detector area information is contained within the response matrix
files (\rmfs), which are \emph{not} normalized to unity.  In practice,
before fitting the pile up model, we renormalize the gratings \fst\
order \arfs\ using the \isis\ functions {\tt factor\_rsp}, {\tt
get\_arf}, and {\tt put\_arf}.  Assuming {\tt a\_id} is the data index of
the \fst\ order \arf, and {\tt r\_id} is the data index of the \fst\
order \rmf, we proceed as follows in \isis:
\begin{verbatim}

   isis> a = get_arf(a_id);        % Read the existing arf into a structure
   isis> na_id = factor_rsp(r_id); % Factor the rmf into normalized rmf*arf
   isis> na = get_arf(na_id);      % Read the new, factored arf component
   isis> a.value *= na.value;      % Multiply original arf by factored value
   isis> put_arf(a_id, na);        % Replace old arf value with rescaled value
   isis> delete_arf(na_id);        % Delete the factored arf component
\end{verbatim}
It is precisely because \isis\ has the ability to
read and manipulate information from all input data files, and then
easily mathematically manipulate this information to create new
functionality, that we are able to define a relatively simple pile up
correction model.

Ideally, we should incorporate additional information, such as the
variation of the line spread function (LSF; i.e., the cross dispersion
profile) along the gratings arms, the 5--10\% of the events that are
typically excluded by order sorting\footnote{LSF variations and the
exclusion of events that fall outside of the order sorting windows is,
of course, accounted for in calculation of the gratings \arfs.}, etc.
For these reasons we cannot assign an absolute estimate of the pile up
fraction.  However, to the extent that we have correctly captured the
\emph{relative} changes of \emph{total} count rate along the
dispersion direction, and have chosen a suitable peak pile up
fraction, we should have a reasonably accurate pile up correction as a
function of wavelength.  More sophisticated {\sl ab initio} models
currently are under development (J. Davis, priv. comm.).

For our data, the peak of the detected \meg\ count rate (including
contributions from $2^{\rm nd}$ and $3^{\rm rd}$ order events) occurs
in the 6--8\,\AA\ region, and peaks at a value of $\approx$ 0.08
counts/frame/3 pixels.  Thus, making a simple first order correction
since this rate already includes pile up, and including up to 5 pixels,
we expect a peak pile up fraction in the 0.09--0.16 range.  We could
have chosen a value in this range and have frozen it in the fits, but
the fitted values fall within the middle of these estimates.
Likewise, for the \heg\ spectra we estimate a peak pile up fraction of
0.025--0.036.  Such fractions are comparable to or smaller than
existing systematic uncertainties in the \chandra\ calibration.
Additionally, we found a great deal of degeneracy in any attempts to
fit the \heg\ value directly, thus we froze peak the pile up fraction at a
value of 0.03.

The {\tt S-lang} code that we used to define a simple gratings pile up
model as an \isis\ user model is presented in the electronic version
of this manuscript.

\begin{verbatim}

define simple_gpile_fit(lo,hi,par,fun)
{  
   % Peak pileup correction goes as:
   %    exp(log(1-pfrac)*[counts/max(counts)])

   variable pfrac = par[0];

   % Pileup scales with model counts from *data set* indx ...
   variable indx = typecast(par[1],Integer_Type);

   if( indx == 0 or pfrac == 0. )
   {
      return fun;   % Quick escape for no changes ...
   }

   % ... but the arf index could be a different number, so get that

   variable arf_indx = get_data_info(indx).arfs;

   % Get arf information

   variable arf = get_arf(arf_indx[0]);

   % In dither regions (or bad pixel areas), counts are down not
   % from lack of area, but lack of exposure.  Pileup fraction
   % therefore should scale with count rate assuming full exposure.
   % Use the arf "fracexpo" column to correct for this effect

   variable fracexpo = get_arf_info(arf_indx[0]).fracexpo;
   if( length(fracexpo) > 1 )
   {
      fracexpo[where(fracexpo ==0)] = 1.;
   }
   else if( fracexpo==0 )
   {
      fracexpo = 1;
   }

   % Rebin arf to input grid, correct for fractional exposure, and 
   % multiply by "fun" to get ("corrected") model counts per bin

   variable mod_cts;
   mod_cts = fun * rebin( lo, hi,
                          arf.bin_lo, arf.bin_hi,
                          arf.value/fracexpo*(arf.bin_hi-arf.bin_lo) )
                / (hi-lo);

   % Go from cts/bin/s -> cts/angstrom/s 

   mod_cts = mod_cts/(hi-lo);

   % Use 2nd and 3rd order arfs to include their contribution.
   % Will probably work best if one chooses a user grid that extends
   % from 1/3 of the minimum wavelength to the maximum, and has
   % at least 3 times the resolution of the first order grid.

   variable mod_ord;
   if(par[2] > 0)
   {
      indx = typecast(par[2],Integer_Type);
      arf = get_arf(indx);
      fracexpo=get_arf_info(indx).fracexpo;
      if( length(fracexpo) > 1 )
      {
         fracexpo[where(fracexpo ==0)] = 1.;
      }
      else if( fracexpo==0 )
      {
         fracexpo = 1;
      }
      mod_ord = arf.value/fracexpo*
                rebin(arf.bin_lo,arf.bin_hi,lo,hi,fun);
      mod_ord = rebin(lo,hi,2*arf.bin_lo,2*arf.bin_hi,mod_ord)/(hi-lo);
      mod_cts = mod_cts+mod_ord;
   }
   if(par[3] > 0)
   {
      indx = typecast(par[3],Integer_Type);
      arf = get_arf(indx);
      fracexpo=get_arf_info(indx).fracexpo;
      if( length(fracexpo) > 1 )
      {
         fracexpo[where(fracexpo ==0)] = 1.;
      }
      else if( fracexpo==0 )
      {
         fracexpo = 1;
      }
      mod_ord = arf.value/fracexpo*
                rebin(arf.bin_lo,arf.bin_hi,lo,hi,fun);
      mod_ord = rebin(lo,hi,3*arf.bin_lo,3*arf.bin_hi,mod_ord)/(hi-lo);
      mod_cts = mod_cts+mod_ord;
   }

   variable max_mod = max(mod_cts);
   if(max_mod <= 0){ max_mod = 1.;}

   % Scale maximum model counts to pfrac

   mod_cts = log(1-pfrac)*mod_cts/max_mod;

   % Return function multiplied by exponential decrease

   fun = exp(mod_cts) * fun;
  
   return fun;
}

add_slang_function("simple_gpile",
                   ["pile_frac","data_indx","arf2_indx","arf3_indx"]);

set_function_category("simple_gpile", ISIS_FUN_OPERATOR);
\end{verbatim}

%\bibliographystyle{jwapjbib}
%\bibliography{mnemonic,jw_abbrv,apj_abbrv,bhc,agn,diplom,inst,ns,conferences}

\end{document}